\documentclass[prc,aps,twocolumn,floatfix]{revtex4}

\usepackage[dvips]{graphicx}
\usepackage{epsfig}
\usepackage{pst-plot}
\usepackage{bm}
\usepackage{amsfonts}

\begin{document}

\title{Effective Interaction Techniques for the Gamow Shell Model}

\author{Gaute Hagen}
\affiliation{Centre of Mathematics for Applications, University of Oslo, 
N-0316 Oslo, Norway}  
\affiliation{Department of Physics and Technology, University of Bergen,
N-5007 Bergen, Norway} 

\author{M.~Hjorth-Jensen}
\affiliation{Department of Physics and Centre of Mathematics for Applications, 
University of Oslo, N-0316 Oslo, Norway}
\affiliation{PH Division, CERN, CH-1211 Geneva 23, Switzerland}
\affiliation{Department of Physics and Astronomy, Michigan State University, 
East Lansing, MI 48824, USA}
\author{Jan S.~Vaagen}
\affiliation{Department of Physics and Technology, University of Bergen, 
 N-5007 Bergen, Norway} 
\date{\today}
\begin{abstract}
We apply a contour deformation technique in momentum space 
to the newly developed Gamow shell model, and study the 
drip-line nuclei ${}^5$He, ${}^6$He and ${}^7$He. 
A major problem in Gamow shell-model studies of nuclear
many-body systems is the increasing dimensionality of many-body 
configurations due to the large number
of resonant and complex continuum states necessary to reproduce 
bound and resonant state energies. We address this problem
using two different effective operator approaches 
generalized to the complex momentum plane.      
These are the Lee-Suzuki similarity transformation method
for complex interactions and the multi-reference
perturbation theory method. The combination of these two approaches results in a large 
truncation of the relevant configurations compared 
with direct diagonalization.
This offers interesting perspectives for studies of weakly bound systems.   
\end{abstract}

\maketitle

\section{Introduction}
\label{sec:introduction}

We expect that present and proposed nuclear structure research facilities
for radioactive beams 
will open new territory into regions of heavier nuclei.
Such systems pose significant challenges to existing 
nuclear structure models since many of
these nuclei will be unstable and short-lived. How to deal with weakly
bound systems and coupling to resonant states is an open and interesting problem in
nuclear spectroscopy. Weakly bound systems cannot be properly  described within a 
standard shell-model approach since even bound states exhibit a strong coupling with
the continuum.

It is therefore important to investigate theoretical methods that will allow
for a description of systems involved in such
element production.  Ideally, we would like to start from an ab initio 
approach  with the free nucleon-nucleon interaction and eventually also 
three-body interactions as the basic building
blocks for the derivation of an effective shell-model interaction. 
The newly developed Gamow shell model offers such a possibility, 
see for example 
Refs.~\cite{michel1,michel2,michel3,liotta,betan,witek1,witek2,roberto,betan2}.
Similarly, the recent work on the continuum shell-model by Volya and Zelevinsky
\cite{vz2003} conveys similar interesting prospectives. Here we focus on the 
Gamow shell model, which has proved 
to be a powerful tool in describing and understanding the formation of 
multi-particle resonances within a shell-model formulation. Representing
the shell-model equations using a  Berggren basis
\cite{berggren,berggren1,berggren2,berggren3,lind,hagen}, allows for a simple
interpretation of multi-particle resonances in terms of single-particle
resonances, as opposed to the traditional harmonic oscillator representation, where 
resonances never appear explicitly. 

Although the Gamow shell-model approach is a powerful tool in this respect, there
are major computational and theoretical challenges that
need to be overcome if we aim at a realistic description 
of weakly bound and unbound nuclei. 
One of the challenges regarding the Gamow shell model 
discussed in Refs.~\cite{liotta, betan}, was
the problem of choosing a contour in the complex $k$-plane that 
in the many-particle case selects the physical interesting
states from the dense distribution of continuum states. 
In Refs.~\cite{liotta,betan}  the authors employ a ``square-well'' contour, which in the 
two-particle case separates the resonances from the complex-continuum 
states. In the case where more than two particles are present in the shell-model space, 
the resonant states mix  with the complex continuum states,
and an identification of the multi-particle resonances becomes difficult.

In this work we consider as a test case the light 
drip-line nuclei $^{5,6,7}$He, and the formation 
of resonances in these nuclei 
starting from a single-particle picture. These nuclei have also been 
studied with a number of other methods, see for example Ref.~\cite{jonson}, 
and references therein.
We  construct a single-particle basis using the contour deformation method 
in momentum space, discussed in detail in Ref.~\cite{hagen}, see also Ref.~\cite{nimrod}
for further references on complex scaling.
We show that choosing a rotated plus translated contour in the complex
plane, a large portion of the many-particle energy surface is free from
complex continuum states. This choice of contour isolates the physical resonances,
and allows for a clear distinction of many-particle resonances 
from the dense distribution 
of complex continuum states, also in the case when the number of particles 
exceeds two. 

The most severe problem and future challenge is that the 
shell-model dimension increases dramatically for $n > 2$
particles moving in a large valence space, 
this is what we henceforth refer to as the dimensionality problem.
Using a technique such as the traditional Lanczos iteration method \cite{lanczo}
fails in Gamow shell-model calculations.
Dealing with large real symmetric matrices, the Lanczos scheme is a powerful 
method when one wishes to calculate the states lowest in energy.
In Gamow shell-model calculations there may be  a large 
number of complex continuum states
lying below the physical resonances in real energy.
In addition it is difficult to
predict where the multi-particle resonances will appear after diagonalization. 
In Refs.~\cite{witek1,witek2} this problem was circumvented by 
choosing a small number of complex continuum states in the single-particle basis. 
It was also pointed out that the results obtained 
were not converged with respect to the number of single-particle continuum orbits.
In Ref.~\cite{michel2} another approach was considered, where at
most two particles where allowed to move in complex continuum states.
This was based on the assumption that these configurations play the dominant
role in the formation of many-particle resonances, and configurations where
more than two particles move in continuum states could be neglected.

Our aim in this work is to propose an effective interaction scheme 
which allows for a much larger number of complex continuum states in the calculations, 
and in addition takes into account the mixing of 
configurations where all particles may move in 
complex continuum states. We show that
if one aims at accurate calculations of the multi-particle resonances,
the effect of all particles moving in the continuum may not always be
neglected. Our choice of contour
allows for a perturbative treatment of the many-particle resonances, 
and we propose a perturbation theory based scheme which combines 
the Lee-Suzuki similarity transformation  
method \cite{suzuki1,suzuki2,suzuki3,suzuki4}
and the so-called  multi-reference perturbation method \cite{multi1,multi2,multi3}
to account for couplings
with configurations where all single-particles move in complex continuum states.

Presently, Gamow shell-model calculations have been performed with phenomenological 
nucleon-nucleon interactions. A major challenge is to construct
effective nucleon-nucleon interactions for drip-line nuclei starting from 
a realistic nucleon-nucleon interaction. 
In this paper we focus on the choice of contour and the dimensionality
problem. The effective nucleon-nucleon interaction adopted is purely
phenomenological. 
However, the scheme we present, although implemented with a phenomenological 
nucleon-nucleon interaction, allows to define effective interactions computed with the 
complex scaled single-particle basis.
The problem of constructing an effective interaction based on present interaction models
for the nucleon-nucleon force will be considered in a forthcoming work.

The outline of this work is as follows. Sec.~\ref{sec:formalism} gives a brief
description of the contour deformation method in momentum space, 
and presents calculations
of the energy spectrum of the nuclei $^{5,6,7}$He. 
Sec.~\ref{sec:suzuki} presents first the 
Lee-Suzuki transformation method generalized to complex interactions. Thereafter we
apply the similarity transformation method to the unbound nucleus $^7$He and give a 
convergence study of its $J^\pi=3/2^-_{1}$ ground state resonance. 
Sec.~\ref{sec:multi} gives a brief
outline of the multi-reference perturbation method, 
and its application to the ground state
of  $^{7}$He. In Sec.~\ref{sec:scheme} we present 
an effective interaction scheme, which combines 
the Lee-Suzuki similarity transformation and the multi-reference perturbation method, 
for calculation of 
multi-particle resonances in weakly bound nuclei. Sec.~\ref{sec:conclusion} 
gives the conclusions of the present study and 
future perspectives and challenges for 
Gamow shell-model calculations.

\section{The Gamow Shell Model}
\label{sec:formalism}

The newly developed Gamow shell model has proved to be a powerful tool in 
describing and understanding multi-particle resonances appearing in 
nuclei near the drip-lines. Here we discuss how two- and three-particle
resonances are formed in $^6$He and $^7$He, and how they are to be understood
in terms of a single-particle picture. The specific choice of contour in 
the complex $k$-plane makes it easy to identify and interpret the multi-particle
resonances. In this section no truncations are made, 
all possible configurations within a model space are used in the shell
model calculations.

\subsection{Berggren Basis in the  Momentum Representation}
In Ref.~\cite{hagen} we studied the contour deformation method applied to 
the momentum space Schr\"odinger equation. It was 
discussed and shown how the specific choices of contours  based on the analytic 
structure of the potential may allow for a unified description of
bound, anti-bound (virtual) and resonant states.
We will apply this method to obtain a single-particle Berggren basis 
for use in Gamow shell-model calculations. Here we briefly outline the 
contour deformation method, and refer the reader to Ref.~\cite{hagen} 
for a more rigorous discussion.

\begin{figure}[hbtp]
\begin{center}
\resizebox{8cm}{5cm}{\epsfig{file=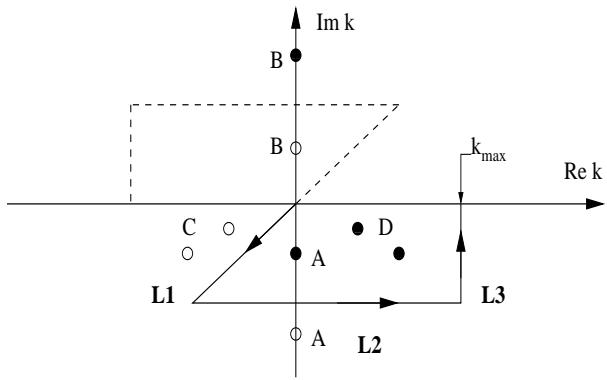}}
\end{center}
\caption{Contour $ C_{R+T}^{+} = L_{1} + L_{2} + L_{3} $ is given 
by the solid line, while
the contour $C_{R+T}^{-} $ is given by the dashed line. 
The contour $C_{R+T} = C_{R+T}^{+}+C_{R+T}^{-}$ is
inversion symmetric. The single-particle spectrum which is 
exposed by this contour is marked by
filled circles
$ \bullet $ and the excluded spectrum by open circles $\circ $. 
The full spectrum includes bound states (B), 
anti-bound (A), decay (D) and capture (C) resonant states.  }
\label{fig:contour2}
\end{figure}

The analytically continued Schr\"odinger equation on a general 
inversion symmetric contour takes the form
\begin{equation}
\label{eq:neweq1}
{\hbar^{2}\over 2\mu}k^2\psi_{nl}(k) + {2\over\pi}\int_{C^{+}} 
dq {q}^{2}V_{l}(k,q)\psi_{nl}(q) = E_{nl}\psi_{nl}(k).
\end{equation}
Here both $k$ and $q$ are defined on an inversion symmetric contour $C^+$ in the lower
half complex $k$-plane, resulting in  a closed integral equation. 
The eigenfunctions constitute a complete bi-orthogonal set, 
normalized according to the Berggren metric \cite{berggren,berggren1,berggren2,berggren3,lind}, namely
\begin{equation}
\label{eq:unity2}
{\bf 1} = \sum _{n\in \bf{C}}\vert\psi_{nl}\rangle\langle\psi_{nl}^{*}\vert + 
\int_{C^{+}} dk k^2\vert\psi_{l}(k)\rangle\langle\psi_{l}^{*}(k)\vert.   
\end{equation} 
In this work we  construct a single-particle Berggren ensemble 
on a rotated plus translated contour, $C_{R+T}$, in the complex $k$-plane, 
studied in detail in Ref.~\cite{hagen}. 
The contour $C_{R+T}^{+}$ is part of the inversion symmetric 
contour $C_{R+T} = C_{R+T}^{+} + 
C_{R+T}^{-}$ displayed in Fig.~\ref{fig:contour2}.
The complete set of single-particle orbits defined by this contour will then include 
anti-bound, bound and resonant states. This basis serves as our starting point 
for Gamow shell-model calculations.

\subsection{Single-Particle Spectrum of $^5$He}
\label{subsec:helium5}
We consider first the unbound nucleus $^5$He. This nucleus may be
modeled by an inert $^4$He core with a neutron moving mainly in the resonant 
spin-orbit partners $p_{3/2}$ and $p_{1/2}$. 
The  $J^\pi={3/2^{-}_1}$ resonance, to be associated with the single-particle orbit 
$p_{3/2}$, is experimentally 
known to have a width of $\Gamma \approx 0.60$ MeV while the 
$J^\pi={1/2^-_1}$ resonance, associated with the single-particle orbit 
$p_{1/2}$, has a large width 
$\Gamma \approx 4$ MeV. For more information on these systems, 
see for example the recent review by Jonson \cite{jonson}.  
The core-neutron interaction in $^5$He may be phenomenologically modeled by the 
SBB (Sack, Biedenharn and Breit) potential ~\cite{SBB}. 
The SBB potential is of Gaussian type with a spin-orbit term, fitted to
reproduce the neutron - $^4$He scattering phase shifts.
In momentum space the SBB potential, which consist of a central part $c$ and 
a spin-orbit term $\vec{\sigma} \vec{l}$, reads
\begin{equation}
   V_{lj}(k,k') = V_{lj}^c(k,k') + ( { \vec{ \sigma}
   \cdot\vec{{ l} } }\:)V_{lj}^{\sigma l}(k,k'),
\end{equation}  
with
\begin{equation}
  V_{l,j}^i(k,k') = -g_i {\pi \over 4\alpha ^2_i}{1\over \sqrt{kk'}}
  \exp ( {-\left( {k^2+ {k'}^2\over 4\alpha ^2_i}\right)} ) 
  I_{l + 1/2}\left( {kk'\over 2\alpha^2_i} \right),
\end{equation}
where the subscripts $lj$ refer to the single-particle orbital and angular momentum
quantum numbers $l$ and $j$, respectively. 
The term $I_{l+1/2}(z) $ is a Bessel function of the first kind with complex
arguments. Fitting this potential to reproduce the $^5$He 
single-particle spectrum and phase-shifts results in
  $g_c=47.4$ MeV, $g_{\sigma l} = 5.86$ MeV and 
$\alpha_c  =  \alpha_{\sigma l} = 2.3$ fm$^{-1}$.

In the complex $k$-plane the Gaussian potential diverges exponentially 
for $\vert \mathrm{Im}[k] \vert > \vert \mathrm{Re}[k]\vert$. 
If we apply the complex scaling technique, which consists of
solving the momentum space Schr\"odinger equation on a purely rotated contour,
we get the restriction $\theta < \pi/4$ on the rotation angle. Even for smaller angles
we may get a poor convergence, since the Gaussian potential 
oscillates strongly along the rotated contours. 
On the other hand, choosing a contour of the type $C_{R+T}^+$ solves this problem, 
allowing for a continuation in the third quadrant of the 
complex $k$-plane. Furthermore, it yields a faster and smoother decay of the 
Gaussian potential along the chosen contour.

\begin{figure}[hbtp]
\begin{center}
\resizebox{8cm}{5cm}{\epsfig{file=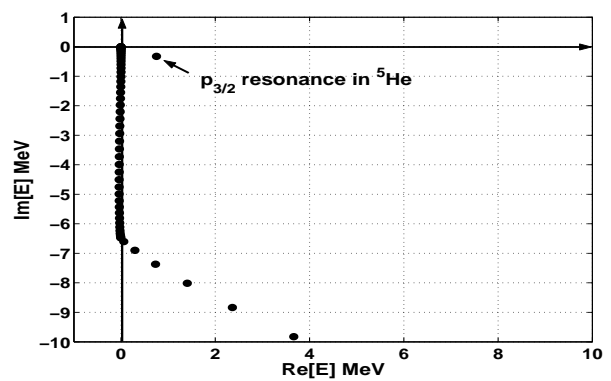}}
\end{center}
\caption{Plot of the $p_{3/2}$ single-particle spectrum in ${}^5$He 
for a Gaussian single-particle potential. The resonance is well located. The remaining points represent the non-resonant continuum.}
\label{fig:fig2}
\end{figure}

\begin{figure}[hbtp]
\begin{center}
\resizebox{8cm}{5cm}{\epsfig{file=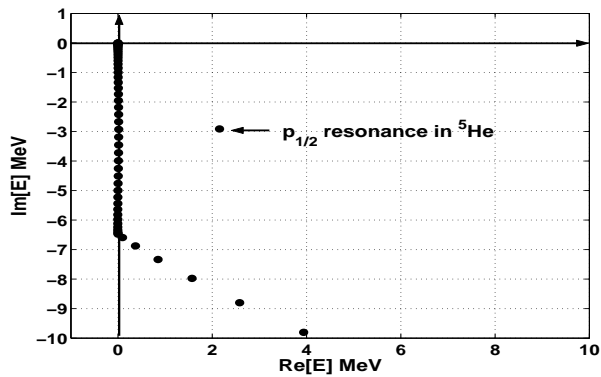}}
\end{center}
\caption{Plot of the $p_{1/2}$ single-particle spectrum in ${}^5$He 
for a Gaussian single-particle potential. The resonance is well located. The remaining points represent the non-resonant continuum.}
\label{fig:fig3}
\end{figure}

Since $^5$He has only resonances in its spectrum, viz., no anti-bound states, 
there is no need for an analytic continuation
in the third quadrant of the complex $k$-plane, as done in Ref.~\cite{hagen} for the 
free nucleon-nucleon interaction. 
We choose a contour of the type $C^+_{R+T}$  rotated with $\theta = \pi/4$ 
and translated with 
$\vert \mathrm{Im}[k] \vert = 0.4 \sin (\pi/4) \approx 0.28$ fm$^{-1}$ in the 
fourth quadrant of the complex $k$-plane.  
Figs.~\ref{fig:fig2} and \ref{fig:fig3} give
plots of the single-particle spectrum in ${}^5$He for the spin-orbit
partners $p_{3/2}$ and $p_{1/2}$ respectively. 
We have used 50 integration points along the rotated $C_R$, and the translated $C_T$ parts of the
contour $C^+_{R+T}$ in the complex $k$-plane. 

Table~\ref{tab:tab1} gives the convergence of the $p_{3/2} $ and the $p_{1/2}$ 
single-particle resonances as function of integration points along the contour 
$C_{R+T}$. 
We observe that with $12$ points along the rotated path 
and $12$ points along the translated
line, one has a reasonable  convergence of the resonance energy, giving in total 
$48$ single-particle states for the valence space consisting of the $lj$ orbits
$\{p_{3/2}, p_{1/2}\}$ with their pertinent momenta $k$ defined by the number of mesh points.
It is clear that if several particles were to move in this space, the dimensionality 
would become enormous. It is therefore important, even at the single-particle 
stage, to  optimize the distribution of continuum states, in order  that 
the main features 
of the system are reproduced with a small number of single-particle resonances and
complex continuum states. 
\begin{table}[htbp]
\caption{Convergence of $p_{3/2}$ and $p_{1/2} $ resonance energies in ${}^5$He as 
function of the number of 
integration points $N_R$ along the rotated $C_R$ and $N_T$ along 
the translated part $C_T$ of the contour. Energies are given in units of MeV.}
\begin{tabular}{rrrrrr}\hline
  \multicolumn{2}{c}{} & 
  \multicolumn{2}{c}{$J^\pi = {3/2}^-$} & 
  \multicolumn{2}{c}{$J^\pi = {1/2}^- $}\\
  \hline
  \multicolumn{1}{c}{$N_R$} & \multicolumn{1}{c}{$N_T$} & 
  \multicolumn{1}{c}{Re[E]} & \multicolumn{1}{c}{Im[E] }& 
  \multicolumn{1}{c}{Re[E]} & \multicolumn{1}{c}{Im[E] }\\
  \hline
  10 & 10 &  0.752321 & -0.329830 &   2.148476 & -2.912522 \\
  12 & 12 &  0.752495 & -0.327963 &   2.152992 & -2.913609 \\
  20 & 20 &  0.752476 & -0.328033 &   2.154139 & -2.912148 \\
  30 & 30 &  0.752476 & -0.328033 &   2.154147 & -2.912162 \\
  40 & 40 &  0.752476 & -0.328033 &   2.154147 & -2.912162 \\
  \hline
\end{tabular}
\label{tab:tab1}
\end{table}
Notice also that the calculated width of the ${1/2}^-$ resonance is somewhat larger
($\approx 6$ MeV) than the experimental  value ($\approx 4$ MeV), see Ref.~\cite{jonson}.

\subsection{Two-Particle Resonances in $^6$He}
Here we present results for  the resonant spectra of $^6$He. We employ again a 
shell-model picture with $^6$He modeled by an inert $^4$He core and
two valence neutrons moving in the $lj$ orbits $\{p_{3/2}, p_{1/2}\}$,
ignoring the recoil from the core.
The model space consists then of all momenta $k$ defined by the set of 
mesh points along the various contours, pertinent to these two $lj$ orbits.
Using the single-particle wave functions for $^5$He of Subsec~\ref{subsec:helium5}, 
we can in turn construct 
an anti-symmetric two-body wave function based on these single-particle wave functions,
viz., 
\begin{equation}
  \Psi_\alpha^{JM} (1,2) = \sum_{a \leq b} C_{a,b}^{JM} {\Phi}_{a,b}^{JM}(1,2), 
\end{equation}
where the indices $a,b$ represent the various single-particle orbits.
Here ${\Phi}_{a,b}^J(1,2) $ is an anti-symmetric two-particle basis state 
in the $j-j$ coupling scheme. The sum over single-particle orbits is limited
by $a\le b$ since we deal with identical particles only.
The expansion coefficients fulfill the completeness relation
\begin{equation}
  1 = \sum_{a\leq b} \left( C_{a,b}^{JM} \right)^2,
\end{equation}
  and the two-particle Berggren basis forms a complete set
\begin{equation}
  1 = \sum_{a\leq b} \vert {\Phi}_{a,b}^{JM}(1,2) \rangle \langle 
\tilde{\Phi}_{a,b}^{JM}(1,2)\vert.  
\end{equation}
Here $\langle \tilde{\Phi}_{a,b}^{JM}(1,2)\vert $ is the complex conjugate of 
$ \langle {\Phi}_{a,b}^{JM}(1,2)\vert $.
As an effective two-neutron interaction $V_{ij} $ we use a 
phenomenological interaction of Gaussian type, separable in $\bf{r}_i,\bf{r}_j$, 
given by 
\begin{equation}
  V_{ij}({\bf{r}_i,\bf{r}_j}) =  V_0 \exp( -\alpha^2(r_i^2 + r_j^2) )
\sum_\lambda ( Y_{\lambda}(i)\cdot Y_{\lambda}(j) ). 
\label{eq:twobody}
\end{equation}
Two model spaces were considered. The first case includes only the  
$p_{3/2}$ single-particle orbit for various values of the momentum $k$ to be defined below.
The second model space includes also the 
$p_{1/2}$ single-particle orbit and its relevant momenta. 
For both model spaces we fit the interaction
strength to reproduce the $0^+$ binding energy in $^6$He.
We have observed that the position of the $2^+$ resonance in $^6$He depends 
on the range $\alpha$  of the Gaussian interaction, even though 
the $0^+$ ground state does not change with $\alpha $. 
Unfortunately it turns out that for larger values of
$\alpha$  the energy fit is better, but the convergence as function of meshpoints is poorer. 
In our calculations we have chosen a value of $\alpha $  which is a compromise between a 
small number of mesh points along the contour
and a reasonable good fit of the resonant energy spectra. 
This demonstrates that 
the two-particle resonant spectrum depends on the radial shape of the interaction and
suggests that we should rather deal  with an effective interaction derived from realistic models
for the nucleon-nucleon interaction.
The parameters used in our calculations are
$V_0 =  -5.315$ MeV for the model space involving only  the $(p_{3/2})$ states and  
$V_0 =-4.549$ MeV for a model space consisting of both single-particle quantum states
$p_{3/2}$ and $p_{1/2}$. We  use 
$\alpha=4.8$ fm$^{-1}$ for both model spaces.

\begin{figure}[hbtp]
\begin{center}
\resizebox{8cm}{6cm}{\epsfig{file=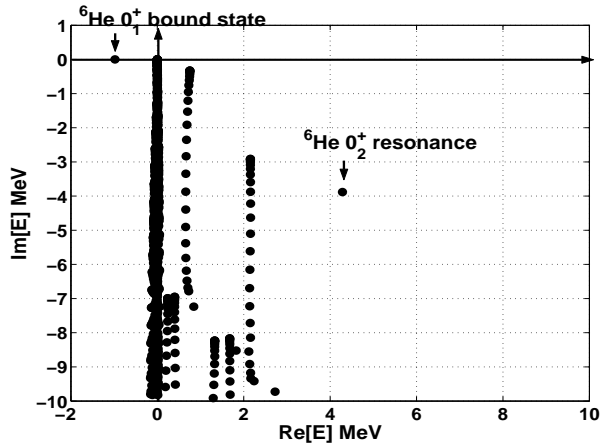}}
\end{center}
\caption{Plot of $0^+_1$ bound- and $0^+_2$ resonant state in $^6$He for a model space consisting of the 
$p_{3/2} $ and 
$p_{1/2} $ single-particle orbits. The bound and resonant states are well located. 
The remaining points represent the non-resonant continuum.} 
\label{fig:he6_0plus}
\end{figure}

\begin{figure}[hbtp]
\begin{center}
\resizebox{8cm}{6cm}{\epsfig{file=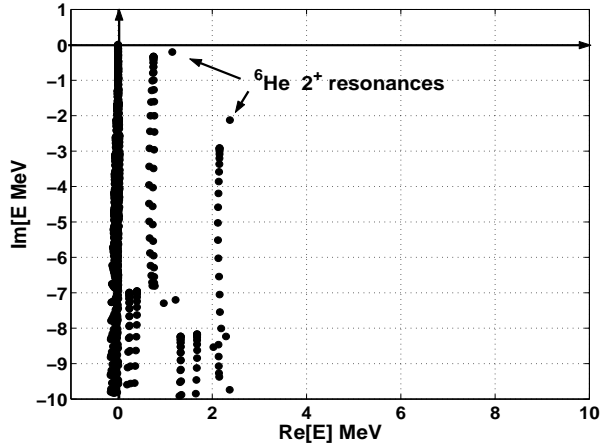}}
\end{center}
\caption{Plot of $2^+$ resonances in $^6$He for a model space consisting of the 
$p_{3/2} $ and 
$p_{1/2} $ single-particle orbits. Both resonant states are well located. 
The remaining points represent the non-resonant continuum.} 
\label{fig:he6_2plus}
\end{figure}

Figs.~\ref{fig:he6_0plus} and \ref{fig:he6_2plus}  show the 
$0^+ $ and $2^+$  energy spectrum, respectively, for $^6$He after a full diagonalization of the
two-particle shell-model equation. Here the model space is defined by the  $p_{3/2}$ and $p_{1/2}$ single-particle 
orbits. This model space yields a bound $0^+$  state as well as a  resonant $0^+$ state. Moreover, we obtain
two resonant $2^+$ states. Observe that the choice of contour 
($C^+_{R+T}$) separates all physical relevant states from the dense distribution of 
complex continuum states in the energy plane. By this choice of contour the 
identification of multi-particle resonances is fairly easy, 
and one may study the resonant trajectories as the interaction strength is varied. 

The stability of the $0^+$ and $2^+$ results as function of the number of mesh points is demonstrated
in Tables \ref{tab:tab2} and \ref{tab:tab3}. 
Limiting first the attention to a model space consisting only of the $p_{3/2}$ orbit,
we note that with $N_R=12$ integration points along the rotated path $C_R$ and $N_T=12$ points
along the translated line $C_T$, convergence is satisfactory, even with a  total of $300$ two-particle states. 
\begin{table}[htbp]
\caption{Convergence of the $0^+_1$ bound state energy in ${}^6$He in terms of the  number  
integration points $N_R$ and $N_T$ along the rotated $C_R$ and the translated part $C_T$ of the contour, respectively. 
The number $N_{2p}$ gives the dimension of the two-particle anti-symmetrized basis. 
Here only $p_{3/2}$ single-particle orbits are included. Energies are given in units of MeV.}
\begin{tabular}{rrrrr}\hline
  \multicolumn{1}{c}{$N_R$} & \multicolumn{1}{c}{$N_T$} & \multicolumn{1}{c}{$N_{2p}$} & \multicolumn{1}{c}{Re[E]} & \multicolumn{1}{c}{Im[E] }\\
  \hline
  12 & 12 & 300 & -0.980067 & -0.000759  \\
  20 & 20 & 820 & -0.979508 &  0.000000 \\
  25 & 25 & 1275& -0.979509 &  0.000000 \\
  \hline
\end{tabular}
\label{tab:tab2}
\end{table}
\begin{table}[htbp]
\caption{Convergence of the $2^+_1$ resonant state energy in $^6$He as function of the number of 
integration points $N_R$ and $N_T$ along the rotated $C_R$ and the translated part $C_T$ of the contour, respectively. 
The number 
$N_{2p}$ gives the dimension of the two-particle anti-symmetrized basis. 
 Here only $p_{3/2}$ single-particle orbits are included, and energies are given in units of MeV.}
\begin{tabular}{rrrrr}\hline
  \multicolumn{1}{c}{$N_R$} & \multicolumn{1}{c}{$N_T$} & \multicolumn{1}{c}{$N_{2p}$} & 
\multicolumn{1}{c}{Re[E]} & \multicolumn{1}{c}{Im[E] }\\
  \hline
  12 & 12 & 300 &  1.215956 & -0.267521 \\
  20 & 20 & 820 &  1.216495 & -0.267745 \\
  25 & 25 & 1275 & 1.216496 & -0.267745 \\
  \hline
\end{tabular}
\label{tab:tab3}
\end{table}
Tables \ref{tab:tab4}, \ref{tab:tab5} and \ref{tab:tab6}  repeat the above 
convergence analysis, 
but now employing a model space consisting of the $p_{3/2} $ and $p_{1/2}$ single-particle orbits,
and including also the results for the lowest-lying $^6$He state with quantum numbers 
$J^{\pi}=1^+$.
Increasing the model space brings several new features. We note in Table \ref{tab:tab4}
that the first excited 
$0^+_2$ state is a resonance. However, the stability of the results as functions of the number of mesh points is
comparable to that seen in Tables \ref{tab:tab2} and \ref{tab:tab3}. With 
approximately $12$ mesh points we obtain results close to the converged ones.
\begin{table}[htbp]
\caption{Convergence of $0^+_1$ bound and the $0^+_2$ resonant state energy in ${}^6$He 
as function of the number of 
integration points $N_R$ and $N_T$ along the rotated $C_R$ and the translated part $C_T$ of the contour, respectively. 
The number 
$N_{2p}$ gives the dimension of the two-particle anti-symmetrized basis. 
Here the $p_{3/2}$ and $p_{1/2}$ single-particle orbits are included. Energies in units of MeV.}
\begin{tabular}{rrrrrrr}\hline
  \multicolumn{3}{c}{} & \multicolumn{2}{c}{$J^\pi = 0^+_1$} & \multicolumn{2}{c}{$J^\pi = 0^+_2$}\\
  \hline
  \multicolumn{1}{c}{$N_R$} & \multicolumn{1}{c}{$N_T$} & \multicolumn{1}{c}{$N_{2p}$} &
  \multicolumn{1}{c}{Re[E]} & \multicolumn{1}{c}{Im[E] }&\multicolumn{1}{c}{Re[E]} & \multicolumn{1}{c}{Im[E] }\\
  \hline
  12 & 12 & 600 &  -0.980111 & -0.000497 &  4.289194 & -3.882119 \\
  20 & 20 & 1640 & -0.979148 & -0.000000 &  4.286186 & -3.882878 \\
  25 & 25 & 2550 & -0.979148 &  0.000000 &  4.286181 & -3.882876 \\
  \hline
\end{tabular}
\label{tab:tab4}
\end{table}
Similar conclusions apply to the $1^+_1$ resonance and the two lowest-lying  
$2^+$ resonant states, see Tables \ref{tab:tab5} and \ref{tab:tab6} for more details.
\begin{table}[htbp]
\caption{Convergence of the $1^+_1$  resonance as function of the number of 
integration points $N_R$ and $N_T$ along the rotated $C_R$ and the translated part $C_T$ of the contour, respectively. 
The number 
$N_{2p}$ gives the dimension of the two-particle anti-symmetrized basis. 
Here the $p_{3/2}$ and $p_{1/2}$ single-particle orbits are included. Energies in units of MeV.}
\begin{tabular}{rrrrr}\hline
  \multicolumn{3}{c}{} & \multicolumn{2}{c}{$J^\pi = 1^+$} \\
  \hline
  \multicolumn{1}{c}{$N_R$} & \multicolumn{1}{c}{$N_T$} & \multicolumn{1}{c}{$N_{2p}$} & 
 \multicolumn{1}{c}{Re[E]} & \multicolumn{1}{c}{Im[E] } \\
  \hline
  12 & 12 & 1128 & 1.945539 & -2.920286 \\
  20 & 20 & 3160 & 1.940263 & -2.930619 \\
  25 & 25 & 4950 & 1.940266 & -2.930608 \\
  \hline
\end{tabular}
\label{tab:tab5}
\end{table}
\begin{table}[htbp]
\caption{Convergence of the $2^+_1$ and $2^+_2 $ resonance energy in $^6$He as function of the number of 
integration points $N_R$ and $N_T$ along the rotated $C_R$ and the translated part $C_T$ of the contour, respectively. 
The number 
$N_{2p}$ gives the dimension of the two-particle anti-symmetrized basis. 
Here the $p_{3/2}$ and $p_{1/2}$ single-particle orbits are included. Energies in units of MeV.}
\begin{tabular}{rrrrrrr}\hline
  \multicolumn{3}{c}{} & \multicolumn{2}{c}{$J^\pi = 2^+_1$} & \multicolumn{2}{c}{$J^\pi = 2^+_2$}\\
  \hline
  \multicolumn{1}{c}{$N_R$} & \multicolumn{1}{c}{$N_T$} & \multicolumn{1}{c}{$N_{2p}$} & 
 \multicolumn{1}{c}{Re[E]} & \multicolumn{1}{c}{Im[E] } & \multicolumn{1}{c}{Re[E]} & \multicolumn{1}{c}{Im[E] }\\
  \hline
  12 & 12 & 876 &   1.149842 & -0.203052 & 2.372295 & -2.122474 \\ 
  20 & 20 & 2420 &  1.150527 & -0.203060 & 2.372818 & -2.123253 \\
  25 & 25 & 3775 &  1.150527 & -0.203060 & 2.372817 & -2.123254 \\
  \hline
\end{tabular}
\label{tab:tab6}
\end{table}
We note that the experimental value for the width of the first excited $J^\pi = 2^+_1$ is $\Gamma \approx 113$ KeV
and the energy is $Re[E]_{2^+_1}=1797$ KeV. Our simplified nucleon-nucleon interaction gives
a qualitative reproduction of the data. In a future work we plan
to include a realistic nucleon-nucleon interaction for studies of such systems.

We end this subsection by analyzing the squared amplitude of the single-particle configurations 
${\vert RR \rangle, \vert RC\rangle, \vert CC\rangle}$ of the $0^+, 1^+$ and $2^+$ 
bound- and resonant wave functions. The results are shown in 
Tables~\ref{tab:tab7},\ref{tab:tab8}, \ref{tab:tab77}, \ref{tab:tab9} and \ref{tab:tab10}. 
The reason for doing this analysis is due to the fact that our single-particle 
basis consists of resonant and continuum single-particle orbits.
By performing such an analysis we can disentangle the 
contribution from for example the non-resonant continuum. 
In these tables, $\vert RR \rangle$ stands for both single-particle orbits being  
a resonant single-particle orbit, 
$\vert RC\rangle$ means that one single-particle orbit is 
a resonant single-particle orbit and the other a non-resonant continuum single-particle orbit, while for
$\vert CC\rangle$ both single-particle orbits are from the non-resonant single-particle continuum.
All the results show 
that the configurations where both single-particles are 
resonant orbits, have the largest amplitude in the two-body wave function. 
It is also seen, that the configurations where both particles are in 
complex continuum states have a small effect on the formation of 
two-particle resonances in $^6$He. This is a useful result which we will exploit below when we define 
effective interactions for smaller spaces.

\begin{table}[htbp]
\caption{Expansion coefficients of the $0^+_1$ bound state in $^6\mathrm{He}$.
 The $p_{3/2}$ and $p_{1/2}$  single-particle orbits define the model space. See text for further discussions.}
\begin{tabular}{rrrrr}
\hline
  \multicolumn{1}{c}{} & \multicolumn{2}{c}{$\left( {p}_{3/2}^2\right)$} & 
  \multicolumn{2}{c}{$\left( {p}_{1/2}^2\right) $} \\ 
\hline
  \multicolumn{1}{c}{} & \multicolumn{1}{c}{Re[$\mathrm{C}^2$]} & \multicolumn{1}{c}{Im[$\mathrm{C}^2$]} & 
\multicolumn{1}{c}{Re[$\mathrm{C}^2$]} & \multicolumn{1}{c}{Im[$\mathrm{C}^2$]} \\ 
  \hline
  $\vert RR \rangle$ &  1.10488 & -0.83161 &  0.22620 & -0.16120 \\
  $\vert RC \rangle$ & -0.06036 &  0.88137 & -0.19842 &  0.22423 \\
  $\vert CC \rangle$ & -0.09716 & -0.04974 &  0.02486 & -0.06305 \\
  \hline
\end{tabular}
\label{tab:tab7}
\end{table}
\begin{table}[htbp]

\caption{Expansion coefficients of the $0^+_2$ resonance in $^6\mathrm{He}$.
 The $p_{3/2}$ and $p_{1/2}$  single-particle orbits define the model space. See text for further discussions.}
\begin{tabular}{rrrrr}
\hline
  \multicolumn{1}{c}{} & \multicolumn{2}{c}{$\left( {p}_{3/2}^2\right) $} & 
  \multicolumn{2}{c}{$\left({p}_{1/2}^2\right)$} \\ 
\hline
  \multicolumn{1}{c}{} & \multicolumn{1}{c}{Re[$\mathrm{C}^2$]} & \multicolumn{1}{c}{Im[$\mathrm{C}^2$]} & 
\multicolumn{1}{c}{Re[$\mathrm{C}^2$]} & \multicolumn{1}{c}{Im[$\mathrm{C}^2$]} \\ 
  \hline
  $\vert RR \rangle$ & -0.01136 & -0.08003 & 0.90189 &  0.33029 \\
  $\vert RC \rangle$ &  0.04282 & -0.03939 & 0.05966 & -0.24478 \\
  $\vert CC \rangle$ &  0.00617 &  0.00494 & 0.00082 &  0.02896 \\
  \hline
\end{tabular}
\label{tab:tab8}
\end{table}

\begin{table}[htbp]
  \caption{Expansion coefficients of the $1^+$ resonance in $^6\mathrm{He}$.
    The $p_{3/2}$ and $p_{1/2}$  single-particle orbits
    define the model space. See text for further discussions.}
  \begin{tabular}{rrrrrrr}
\hline
\multicolumn{1}{c}{} & \multicolumn{2}{c}{ $\left( p_{1/2}{p}_{3/2}\right)$} & 
  \multicolumn{2}{c}{$\left({p}_{1/2}^2\right)$} & 
  \multicolumn{2}{c}{$\left({p}_{3/2}^2\right)$} \\ 
\hline
  \multicolumn{1}{c}{} & \multicolumn{1}{c}{Re[$\mathrm{C}^2$]} & \multicolumn{1}{c}{Im[$\mathrm{C}^2$]} & 
  \multicolumn{1}{c}{Re[$\mathrm{C}^2$]} & \multicolumn{1}{c}{Im[$\mathrm{C}^2$]} & 
  \multicolumn{1}{c}{Re[$\mathrm{C}^2$]} & \multicolumn{1}{c}{Im[$\mathrm{C}^2$]} \\ 
  \hline
  $\vert RR \rangle$ & 0.71068 & -0.03739 &  &  &  &   \\
  $\vert RC \rangle$ & 0.00381 & -0.06948 & 0.00016 & 0.00003 & 0.02224  & 0.01171  \\
  $\vert CR \rangle$ & 0.20647 &  0.05208 & 0.00037 & 0.00071 & 0.07067  & 0.03807 \\
  $\vert CC \rangle$ &-0.00984 & -0.00149 &-0.00031 &-0.00009 &-0.00424  & 0.00585 \\
  \hline
\end{tabular}
\label{tab:tab77}
\end{table}

\begin{table}[htbp]
\caption{Expansion coefficients of the $2^+_1$ resonance in $^6\mathrm{He}$.
 The $p_{3/2}$ and $p_{1/2}$  single-particle orbits define the model space. See text for further discussions.}
\begin{tabular}{rrrrr}
\hline
\multicolumn{1}{c}{} & \multicolumn{2}{c}{$\left( p_{1/2}{p}_{3/2}\right)$} & \multicolumn{2}{c}{$\left( {p}_{3/2}^2\right)$} \\ 
\hline
\multicolumn{1}{c}{} & \multicolumn{1}{c}{Re[$\mathrm{C}^2$]} & \multicolumn{1}{c}{Im[$\mathrm{C}^2$]} & 
\multicolumn{1}{c}{Re[$\mathrm{C}^2$]} & \multicolumn{1}{c}{Im[$\mathrm{C}^2$]} \\ 
\hline
$\vert RR \rangle$  &  0.11394 & -0.00494&  0.96962 &  0.05539 \\
$\vert RC \rangle$  & -0.00474 &  0.02531& -0.00178 & -0.00018 \\
$\vert CR \rangle $ & -0.02776 & -0.03796& -0.05069 & -0.02708 \\
$\vert CC \rangle$ &   0.00229 & -0.00772& -0.00089 & -0.00282 \\
  \hline
\end{tabular}
\label{tab:tab9}
\end{table}

\begin{table}[htbp]
\caption{Expansion coefficients of the $2^+_2$ resonance in $^6\mathrm{He}$.
 The $p_{3/2}$ and $p_{1/2}$  single-particle orbits define the model space. See text for further discussions.}
\begin{tabular}{rrrrr}
\hline
  \multicolumn{1}{c}{} & \multicolumn{2}{c}{$\left({p}_{1/2}p_{3/2}\right)$}
  & \multicolumn{2}{c}{$\left( {p}_{3/2}^2\right)$}  \\ 
\hline
  \multicolumn{1}{c}{} & \multicolumn{1}{c}{Re[$\mathrm{C}^2$]} & \multicolumn{1}{c}{Im[$\mathrm{C}^2$]} & 
\multicolumn{1}{c}{Re[$\mathrm{C}^2$]} & \multicolumn{1}{c}{Im[$\mathrm{C}^2$]} \\ 
  \hline
  $\vert RR \rangle$  & 0.88847 & -0.03742 &  0.08911 & -0.03742 \\ 
  $\vert RC \rangle$ & -0.04888 &  0.05674 & -0.00104 & -0.00055 \\ 
  $\vert CR \rangle$ &  0.06058 & -0.03336 & -0.00072 & -0.02469 \\
  $\vert CC \rangle$ &  0.01131 & -0.00447 &  0.00115 & -0.00220 \\
  \hline
\end{tabular}
\label{tab:tab10}
\end{table}

\subsection{Three-Particle Resonances in $^7$He}
Finally we consider the unbound nucleus $^7$He, whose ground state ($J^\pi={3/2}^-$) is 
located $\approx 0.5$ MeV above the $^6$He ground state, with a measured width   
$\Gamma \approx 160$ keV. Other continuum structures, with tentative spin assignments 
$J^{\pi} = 1/2^-$, and $J^{\pi} = 5/2^-$, have been observed, see for example  Ref.~\cite{jonson} 
for an extensive review of the experimental situation. 
In this subsection we limit the attention to 
a model defined by the  $p_{3/2} $ single-particle orbits only. 
Thus our single-particle basis from
$^5$He implies that
only a $J^\pi={3/2}^-$ resonance  may be formed. 
The reason we do not include 
the $p_{1/2}$ single-particle orbits is that we aim at a diagonalization  in the
full space, taking into account all complex continuum couplings.
This model calculation will serve as a later reference.
In the case of $24$ mesh points in momentum space 
for the $p_{3/2}$ single-particle quantum numbers $lj$,   
the total dimension $d$ of the ($J^\pi={3/2}^-$) three-particle problem is $d=9224$, 
if, in addition, we were to include $24$ single-particle momenta for the $p_{1/2}$ 
single-particle quantum numbers $lj$, 
we would have roughly  $d\sim 40000$ three-body configurations. 

In Refs.~\cite{witek1} and \cite{witek2}
the dimensionality problem was circumvented by choosing a small number complex 
continuum states, typically of the order of five,
although it was found that a larger number continuum states had to be included to obtain 
converged results.
Our aim in this subsection is to study the full effect of a coupling to the complex
continuum, and show that if one is to obtain accurate results, the effect of all 
particles moving in the continuum on the formation of multi-particle resonances may  
be important. In addition we include a larger number of continuum states, in order to 
achieve  a satisfactory convergence.
As for $^6$He, we construct a three-body wave function 
using the single-particle wave functions
defined in $^5$He. 
The three-body wave function is expanded in a 
three-particle anti-symmetric Berggren basis
\begin{equation}
  \nonumber 
  \Psi^{JM}_\alpha(1,2,3) = \sum _{a\leq b\leq c} C^{JM}_{(a,b)c}{\Phi}^{JM}_{(a,b)c}(1,2,3), 
\end{equation}
where the completeness relation reads
\begin{equation}
  \nonumber 
  1 = \sum _{a\leq b\leq c} \vert {\Phi}^{JM}_{(a,b)c}(1,2,3)\rangle\langle \tilde{\Phi}^{JM}_{(a,b)c}(1,2,3)\vert, 
\end{equation}
with
\begin{equation}
  \nonumber
  1 = \sum _{a\leq b\leq c} ( C^{JM}_{(a,b)c)} )^2 .
\end{equation}

\begin{figure}[hbtp]
\begin{center}
\resizebox{8cm}{6cm}{\epsfig{file=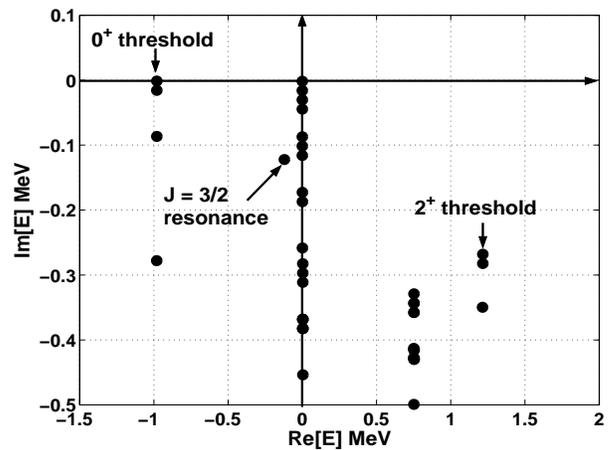}}
\end{center}
\caption{Plot of the ${3/2}^-$ complex energy spectrum of $^7$He 
for a model space consisting of 
$p_{3/2} $ single-particle orbits only. The $J^{\pi}={3/2}^-$ resonance is 
located at $E_{3/2^-}= -(0.120731 +0.122211i)$ MeV. } 
\label{fig:he7_3minus}
\end{figure}

The two-body nucleon-nucleon interaction is the same as that used for $^6$He.
Fig.~\ref{fig:he7_3minus} gives the energy spectrum 
after a full diagonalization of the three-particle shell-model
equation. It is seen that the choice of contour in calculating
the single-particle spectrum is again optimal if we wish to consider the fact that
all physical interesting states are well separated from the dense 
distribution of complex scattering states. 
The $J^{\pi} = 3/2^-$ resonance appears at 
the energy  $E_{3/2^-}= -(0.12 +0.12i)$ MeV.

 The plotted energy spectrum shows that the $0^+$ and $2^+$ states in $^6$He,
 and the ${3/2}^-$ state in $^5$He, form
 complex thresholds in the energy spectrum of the $J^{\pi}=3/2^-$ spectrum in $^7$He,
 see Tables.~\ref{tab:tab1}, \ref{tab:tab2} and \ref{tab:tab3}.
 The physical interpretation of these three-particle states is,
 in the case of the $^6$He thresholds, that
two of the neutrons form either the $0^+$ ground state or the $2^+$ resonant state, 
while the third neutron is moving in a complex continuum state. 
A diagonalization within the reduced space, where at most two particles
move in continuum states gives the resonance energy $-(0.14 +0.16i)$ MeV, 
which shows that the effect coming from all particles moving in the
continuum is not negligible, but small.
 
Table~\ref{tab:tab11} gives the squared amplitudes of the various 
single-particle configurations in the $^7$He ground state, $\{ \vert RRR\rangle, 
\vert RRC\rangle, \vert RCC\rangle, \vert CCC\rangle\} $ , where $R$ again labels 
a single-particle resonance and $C$ a complex single-particle continuum orbit. It is seen 
that the most important configuration, as in the case of $^6$He, 
is the one where all single-particles are
in the $p_{3/2}$ single-particle resonant orbit. The effect of configurations
where all particles are in continuum states is small, which suggest that the
coupling to configurations $\vert CCC \rangle$ may be taken into account
perturbatively.
This a feature we will exploit in Secs.~\ref{sec:suzuki} and \ref{sec:multi}.

In Fig.~\ref{fig:he7_3minus} we note that
the $J^{\pi}=3/2^-$ ground state in $^7$He appears at energy of approximately 
$0.86 $ MeV
above the ground state in $^6$He, while the experimental value is at approximately 
$0.5$ MeV.  
This discrepancy with experiment 
can be understood in terms of the configuration $\vert RRR\rangle$, and the choice of
interaction. Focusing on the first aspect and  
using coefficients of fractional parentage, we can rewrite  
the $\vert RRR\rangle$ configuration as
\begin{equation}
  \nonumber
  \vert (p_{3/2})^3; \: J^{\pi}=3/2^-\rangle = {1\over 6}\vert (p_{3/2})^2_0\: p_{3/2};\:J^{\pi}=3/2^-\rangle - 
\sqrt{5\over 6}\vert (p_{3/2})^2_2 \:p_{3/2};\:J^{\pi}=3/2^-\rangle.
  \label{eq:cfp}
\end{equation}
From the geometry one may conclude that the ground state of $^7$He 
bears much more resemblance with the $2^+_1$ resonance 
than with the $0^+_1$ ground state
of $^6$He. In our calculations the $2^+_1$ resonance comes at an energy  
$ \approx (1.2 -0.26i)$ MeV, which is roughly $2.2$ MeV 
above the $0^+_1$ ground state of $^6$He, 
to be contrasted with  the experimental value 
of $\approx 1.8$ MeV, see Fig.~\ref{fig:he7_3minus}. 
This suggests that if we were
to increase the attractive strength of the $J^{\pi}=2^+$ interaction 
in $^6$He and get a better agreement 
with the experimental value, the $J^{\pi}=3/2^-$ resonant ground state of $^7$He 
would get closer 
to the experimental results.
\begin{table}[htbp]
\caption{Expansion coefficients of the $J^{\pi}={3/2}^-$ ground state in $^7\mathrm{He}$.
 Here only $p_{3/2}$  single-particle orbits are included.}
\begin{tabular}{rrr}
\hline
  \multicolumn{3}{c}{$\vert{p}_{3/2}^3\rangle$}\\
\hline
  \multicolumn{1}{c}{} & \multicolumn{1}{c}{Re[$\mathrm{C}^2$]} & \multicolumn{1}{c}{Im[$\mathrm{C}^2$]} \\
  \hline
  $\vert RRR\rangle$ &  1.295549 & -0.986836 \\
  $\vert RRC\rangle$ & -0.184544 &  1.099729 \\
  $\vert RCC\rangle$ & -0.115738 & -0.110375 \\
  $\vert CCC\rangle$ &  0.004733 & -0.002518 \\
  \hline
\end{tabular}
\label{tab:tab11}
\end{table}

\section{Effective Interactions for the Gamow Shell Model}
\label{sec:suzuki}
\subsection{The Lee-Suzuki Similarity Transformation for Complex Interactions}

The previous section served to introduce and motivate the application of complex scaling
in studies of weakly bound nuclear systems. However, employing such a momentum space basis soon exceeds 
feasible dimensionalities in shell-model studies. 
To circumvent this problem and to be able to define
effective interactions of practical use in shell-model calculations, 
we introduce effective two-body interactions
based on similarity transformation methods. These interactions are in 
turn employed in Gamow shell-model calculations. 
We base our approach on the extensive works of Suzuki, Okamoto, Lee and collaborators, see for example
Refs.~\cite{suzuki1,suzuki2,suzuki3,suzuki4}.
This similarity transformation method has been widely used in the
construction of a effective two- and three-body interactions for use in
the No-Core shell-model approach of Barrett, Navratil, Vary and collaborators, see for 
example Refs.~\cite{bruce1,bruce2,bruce3,bruce4}  
and references therein. However, since  
the similarity transformation method has previously 
only been considered for real interactions, we need to extend
its use to Gamow shell-model calculations, implying a generalization 
to complex interactions. 
To achieve the latter
we introduce first the two-body Schr\"odinger equation
\begin{equation} 
(H_0 + V_{12})\vert\psi_n\rangle = E_n\vert\psi_n\rangle,
\end{equation}
here $H_0$ includes the single-particle part of the Hamiltonian, kinetic energy and 
an eventual single-particle interaction. The term $V_{12}$ is the residual two-body interaction. 
We then expand the exact wave function $\psi_n$ in the anti-symmetric two-particle
basis, generated by the single-particle basis of $H_0$, which corresponds to the basis
from the $^5$He calculations of Subsec.~\ref{subsec:helium5}. 
Thereafter we choose a suitable single-particle model space $p$ and its complement space $q$. 
These single-particle spaces
define in turn our two- and many-particle model spaces 
\begin{equation} 
{P} = \sum_{\alpha_{P}} \vert \alpha_{P}\rangle
\langle \tilde{\alpha}_{P}\vert, 
 \end{equation}
and the complement space 
\begin{equation} 
{Q} = \sum_{\alpha_{Q}} \vert \alpha_{Q}\rangle
\langle \tilde{\alpha}_{Q}\vert, 
\end{equation}
where  $P$ is defined by both 
single-particle orbits being in the $p-$space, and the complement 
space $Q$ is given by all two particle states where at least 
one particle is in the $q$-space.  
The anti-symmetric two-particle basis follows the Berggren metric
\begin{equation} 
\langle \tilde{\alpha'}\vert \alpha\rangle = 
\langle {\alpha'}^*\vert \alpha\rangle = 
\delta_{\alpha',\alpha}, 
\end{equation}
and the projection operators fulfill the relations 
\begin{equation}
P^2 = P, \;\; Q^2 = Q, \;\; P^{\mathrm{T}} = P,
\end{equation}
and 
\begin{equation}
Q^{\mathrm{T}} = Q, \;\; P + Q = 1, \;\; PQ = 0.
\end{equation}
We wish to construct an effective two-body interaction within 
the model space, reproducing in the $P$-space 
exactly the $N_P$ model space eigenvalues of the 
full Hamiltonian. This can be accomplished by a similarity 
transformation
\begin{equation} 
\tilde{H} = e^{-\omega} H e^\omega, 
\end{equation} 
where $\omega$ is defined by $\omega = Q\omega P$. It follows 
that $\omega ^2 = \omega^3 =...=0$ and $e^\omega = P + Q + \omega $.    
The two-body Schr\"odinger equation can then be rewritten in terms of 
a $2\times 2$ block structure 
\begin{equation}
\left( \begin{array}{cc}
P\tilde{H}P & P\tilde{H}Q \\ 
Q\tilde{H}P & Q\tilde{H}Q 
\end{array} \right) 
\left( \begin{array}{c} 
P\psi_n \\
Q\psi_n 
\end{array} \right) = 
E_n \left( \begin{array}{c} 
P\psi_n \\
Q\psi_n 
\end{array} \right).
\end{equation}
If $P\tilde{H}P$ is to be the two-particle effective interaction, the decoupling condition 
$ P\tilde{H}Q = 0$ must be fulfilled. One may show that the decoupling 
condition becomes 
\begin{equation} 
QHP + QHQ\omega - \omega PHP - \omega PHQ\omega = 0,
\label{eq:decoupling}
\end{equation}
with $ \omega  $ acting as a transformation from the model space $P$ to its complement $Q$, viz., 
\begin{equation}
  \nonumber
  \langle\tilde{\alpha}_Q\vert \psi_n \rangle = 
  \sum_{\alpha_P}\langle \tilde{\alpha}_Q\vert\omega\vert\alpha_P\rangle
  \langle\tilde{\alpha}_P\vert \psi_n\rangle.
  \label{eq:suz1}
\end{equation}
There is however no unique solution for $\omega $.
The effective interaction generated in the model space depends on the $N_P$ 
exact solutions entering Eq.~(\ref{eq:suz1}). This is why the effective interaction
generated by the similarity transformation method is often referred to as a 
state dependent effective interaction. 
The solution for $\omega$  may be obtained as long as the matrix 
$ \langle\tilde{\alpha}_P\vert \psi_n\rangle  $ is invertible and non-singular.
Based on this we choose those $N_P$ exact solutions with the largest overlap with 
the two-particle model space. With the solution $\omega $, 
the non-symmetric effective interaction $R$ is given by
\begin{equation} 
  R = P\tilde{H}P - PH_0P = PV_{12}P + PV_{12}Q\omega. 
\end{equation}
It would be preferable to obtain a complex symmetric effective interaction, 
in order to take advantage of the anti-symmetrization of the two-particle basis.
This may be accomplished by a complex orthogonal transformation
\begin{equation}
    V_{\mathrm{eff}} = U^{-1} ( H_0+V_{12} ) U - H_0,
\end{equation}
where $U$ is complex orthogonal and defined by
\begin{equation}
  U = \exp(-S), \:\: S = \mathrm{arctanh}( \omega - \omega^T ),
\end{equation}
and
\begin{equation}
  U^TU = UU^T = 1, \:\: U^T = U^{-1}.
\end{equation}
Such complex orthogonal transformations preserve the Berggren metric $x^Tx$  of any
vector $x \in \{C^{n}\}$. 
This feature allows us to define a complex and symmetric effective two-body interaction
\begin{equation} 
  V_{\mathrm{eff}} = ( P + \omega^{\mathrm{T}}\omega )^{1/2}( PHP + PHQ\omega )( P + \omega^{\mathrm{T}}\omega )^{-1/2} -H_0.
\end{equation}
In determining $V_{\mathrm{eff}}$, one has to 
find the square root of the matrix $A = ( P + \omega ^{\mathrm{T}}\omega )$.  
In the case of $A$ being real and positive definite the method based on
eigenvector decomposition gives generally a stable solution.
Using the eigenvector decomposition, with $Z$ representing an orthogonal matrix,  
$D$ being a diagonal matrix composed of the eigenvalues, using
$A  =  Z D Z^{T}$, $Z^{T}Z = ZZ^{T} = 1$, $D=(D)^{1/2}(D)^{1/2}$, and 
\begin{equation}
  A  =  \left( Z D^{1/2} Z^{T}\right)\left( Z D^{1/2} Z^{T}\right),
\end{equation}
we can write the square root of a matrix $A$ as 
\begin{equation}
  A^{1/2}  =  Z D^{1/2} Z^{T}.  
\end{equation}
For a complex  matrix $A$
the procedure based on eigenvector decomposition is 
generally  numerically unstable. An approach suitable for complex matrices
is based on properties of the matrix sign function. 
It can be shown that the square root of the matrix is 
related to the matrix sign function, 
see Ref.~\cite{higham} for more details.
In the case of $A$ being complex and having all eigenvalues in the open 
right half complex plane, 
iterations based on the matrix sign function are generally more stable
\begin{equation}
  \nonumber
  \mathrm{sign} \left( \left[ \begin{array}{cc} 
      0 & A \\
      I & 0 
    \end{array} \right] \right) = 
  \left[ \begin{array}{cc} 
      0 & A^{1/2} \\ 
      A^{-1/2} & 0  \end{array} \right]. 
\end{equation}
One stable iteration scheme for the matrix sign was derived by 
Denman and Beavers \cite{denman}, as a special case of a method for solving the 
algebraic Riccati equation
\begin{eqnarray}
  Y_0 & = &  A,\:\: Z_0 = I, \\
  Y_{k+1} & = &  {1\over2} ( Y_k + Z_k^{-1}), \\ 
  Z_{k+1} & = & {1\over2} ( Z_k + Y_k^{-1}), \:\: k = 0,1,2,..., 
\end{eqnarray}
and provided $A$ has no non-positive eigenvalues 
this iteration scheme exhibits a quadratic 
convergence rate with
\begin{equation}
  Y_k \rightarrow A^{1/2}, \:\:\: Z_k \rightarrow A^{-1/2} \:\:\: \mathrm{as} \:\:\:k\rightarrow \infty. 
\end{equation}
In our calculations, convergence is typically obtained 
after a few iterations. 

\subsection{Gamow Shell-Model Studies of $^7$He with the Similarity Transformation Method}
\label{subsec:lsresults}
Here we apply the Lee-Suzuki similarity transformation method to the 
Gamow shell-model calculation of the ground state of $^7$He. The first problem
is to define an optimal single-particle model space, which subsequently defines
the two-particle model space, where the effective interaction is constructed. In 
Sec.~\ref{sec:formalism} it was shown that the $J^{\pi}=3/2^-$ 
ground state of $^7$He
has the $2^+$ resonance in $^6$He as an important two-body configuration, 
see Eq.~(\ref{eq:cfp}).
Based on this result, a viable starting point is to study the 
single-particle strengths in the
$2^+$ resonance wave function. To understand the nature of two-particle resonances 
and how they are formed in a shell-model framework, it is natural to study and analyze the single-particle
strengths in the two-particle wave function, and how they are distributed among the single-particle 
resonances and the various complex continuum orbits, given on a specific contour in the
complex $k$-plane. The single-particle density operator is given by
\begin{equation}
  \hat{n}_i = \sum_{j}^N \vert {\psi}_{i}(j)\rangle\langle \tilde{\psi}_{i}(j)\vert, \:\:  
N = \sum_i \hat{n}_i,
\end{equation}
where $N$ is the total number of particles, $i = \{l_i,j_i\}$ 
labels the single-particle quantum numbers and 
$i$ represents the single-particle orbit.
In the case of $^6$He with an inert $^4$He core, $N = 2$. Finding the probability, $n_i$, that 
either particle $1$ or particle $2$ is in the single-particle orbit $i = \{l_i,j_i\}$, 
we calculate the matrix element of $\hat{n}_i$ with the two-particle resonance wave function
\begin{widetext}
\begin{eqnarray} 
  n_i  &=&  \langle \tilde{\Psi}_\alpha^J(1,2)\vert\hat{n}_i\vert {\Psi}_\alpha^J(1,2) \rangle=  \sum_{a\leq b}\sum_{c \leq d} C_{a,b}^\alpha C_{c,d}^{\alpha} \left( {1\over (1+\delta_{a,b}) (1+\delta_{c,d})} \right)^{1/2}\times \nonumber \\
&& 
  \{\delta_{d,i}\:\delta_{c,i}\:\delta_{b,d}+(-1)^{j_c+j_i-J+1}\delta_{a,i}\delta_{d,i}\delta_{b,c}
  +  \delta_{b,i}\delta_{d,i}\delta_{a,c} + (-1)^{j_a+j_i-J+1}\delta_{b,i}\delta_{c,i}\delta_{a,d}\}.
\end{eqnarray} 
\end{widetext}
\begin{figure}[hbtp]
\begin{center}
\resizebox{8cm}{6cm}{\epsfig{file=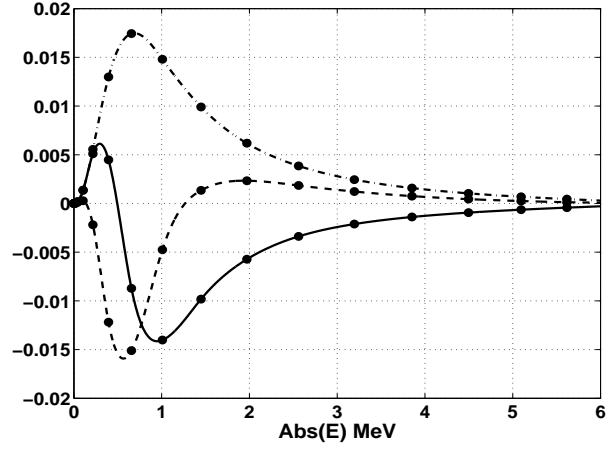}}
\end{center}
\caption{Plot of the $p_{3/2}$ single-particle complex continuum strength $n_i$ in the two-particle
  resonance wave function $2^+_1$ in $^6$He.  The solid line gives the real part, 
  the dotted line the imaginary and the dash-dotted line the absolute value of the
  strengths. The filled circles give the actual location of the complex 
  continuum states in absolute value of energy.} 
\label{fig:sp_2plus}
\end{figure} 
Fig.~\ref{fig:sp_2plus} gives the real, imaginary and absolute values of the single-particle
strengths among the complex continuum orbits, in the $2^+$ resonance in $^6$He. 
The strengths are plotted as a function
of the absolute value of the complex continuum energy. Observe that the continuum states near the 
$2^+$ resonance in $^6$He have the largest strengths. This may be understood as an
interference effect between the single-particle resonances and the continuum 
orbits located closest in energy (momentum) to the single-particle resonance. 
When defining a single-particle model space, we choose the single-particle 
resonant and complex continuum orbits with the largest absolute value of the
single-particle strength. With this recipe we have a consistent way of defining 
a single-particle model space, which forms the basis for constructing 
an effective interaction in the two-particle model space.

\begin{figure}[hbtp]
\begin{center}
\resizebox{8cm}{6cm}{\epsfig{file=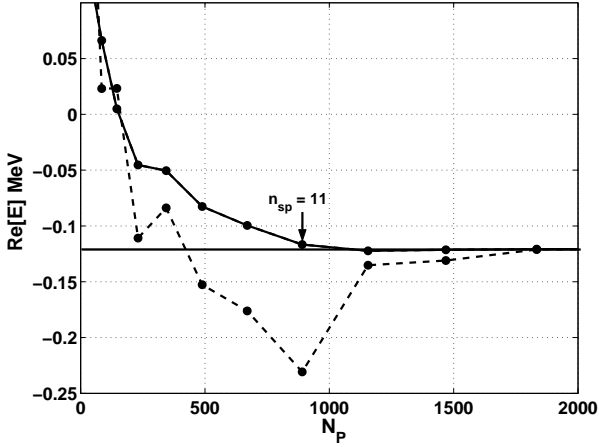}}
\end{center}
\caption{Convergence of the real part of the $J^{\pi}={3/2}^-$ resonance in  $^7$He 
for a space defined by the $p_{3/2} $ single-particle orbits only. The abscissa represents the 
number of three-particle 
model space configurations  $N_P$ while $n_{sp}$ represents the total number of single-particle momenta for the 
$p_{3/2} $ single-particle quantum numbers $lj$. The solid line represents the effective interaction generated by the
Lee-Suzuki similarity transformation method, and the dashed line is obtained using the  the bare interaction
and the same number of three-body configurations. 
The ${3/2}^-$ resonance is 
located at E $= -(0.120731 +0.122211i)$ MeV. The horizontal line is the real energy obtained in the full space of 
three-body configurations.} 
\label{fig:he7_pert1}
\end{figure} 

\begin{figure}[hbtp]
\begin{center}
\resizebox{8cm}{6cm}{\epsfig{file=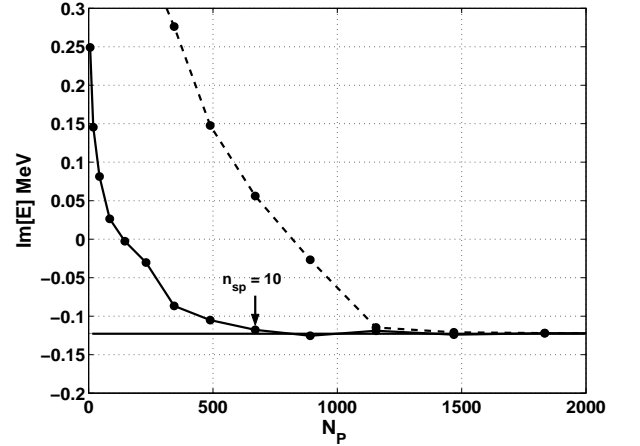}}
\end{center}
\caption{Convergence of imaginary part of $J^{\pi}={3/2}^-$ resonance in  $^7$He 
number of three-particle 
model space configurations$N_P$ while $n_{sp}$ represents the total number of single-particle momenta for the 
$p_{3/2} $ single-particle quantum numbers $lj$. 
The solid line represents the effective interaction generated by the
Lee-Suzuki similarity transformation method, and the dashed line is obtained using the  the bare interaction
and the same number of three-body configurations. 
The $J^{\pi}={3/2}^-$ resonance is 
located at E $= -(0.120731 +0.122211i)$ MeV. The horizontal line is the imaginary energy obtained in the full space of 
three-body configurations.} 
\label{fig:he7_pert2}
\end{figure} 

In Figs.~\ref{fig:he7_pert1} and \ref{fig:he7_pert2} we show the convergence of the 
real and imaginary part of the $J^{\pi}=3/2^-$ resonance in $^7$He, as function of an increasing  
single-particle model space. For comparison we plot the results for 
a diagonalization within the model space using the ``bare'' interaction. 
It is seen 
that results with the effective interaction constructed with the 
similarity transformation method converges much faster
than results obtained with the ``bare'' interaction. 
We see that a satisfactory convergence
is obtained with $10-11$ single-particle Berggren states in the single-particle
model space $p$ from $^5$He, 
corresponding to 
$\approx 700-800 $ three-particle states $N_P$. Compared with the full dimension 
of the three-particle
problem, $9224$, we have effectively reduced the dimension to 
about $8\%$ of the full space. This is a considerable benefit which may allow us to extend the Gamow shell 
 model with a complex scaled single-particle basis to heavier systems and realistic effective interactions.
However, we can further improve upon this approach by considering perturbative techniques as well. That is the topic
of the next Section.

\section{The Multi-Reference Perturbation Method}
\label{sec:multi}
The M\"oller-Plesset multi-reference perturbation method has recently 
been revived in 
quantum chemistry, see for example 
Refs.~\cite{multi1,multi2,multi3}, with an emphasis on scattering theory and 
electron decays in many-body systems.
Here we only give a brief outline of the method, and refer the reader to 
Refs.~\cite{multi1,multi2,multi3} for further details. 

The basic idea of the multi-reference perturbation method is 
to first diagonalize within a small space (reference space), and then 
add a perturbation to the reference states by taking into account excitations
from the reference space to the complement space.   

First we define a suitable $N$-particle reference (model) space 
$P$ which describes most of the many-body correlations of the system, 
and hopefully gives a weak coupling with the complement space $Q$. The $N$-body 
problem may then be written as a  block structure 
\begin{equation} 
  \nonumber
  \left( \begin{array}{cc}
    H^{PP} & H^{PQ} \\ 
    H^{QP} & H^{QQ}  
  \end{array} \right) 
  \left( \begin{array}{c} 
    P\psi_n \\
    Q\psi_n 
  \end{array} \right) = 
  E_n \left( \begin{array}{c} 
    P\psi_n \\
    Q\psi_n 
  \end{array} \right).
\end{equation}
Thereafter we divide the full Hamiltonian in two parts
\begin{widetext} 
\begin{equation} 
  \left( \begin{array}{cc}
    H^{PP} & H^{PQ} \\ 
    H^{QP} & H^{QQ}  
  \end{array} \right)  =  
  \left( \begin{array}{cc}
    H^{PP} & 0 \\ 
    0 & D^{QQ}  
  \end{array} \right) 
  +
  \left( \begin{array}{cc}
    0 & H^{PQ} \\ 
    H^{QP} & \tilde{H}^{QQ}  
  \end{array} \right) =  H^0 + H^1.
\end{equation}
\end{widetext}
Here $ D^{QQ} $ is the diagonal part and $\tilde{H}^{QQ} $ the off-diagonal part of $H^{QQ} $. 
In this form, we see that $H^0$ defines the unperturbed part while $H^1$ gives the 
perturbations to $H^0$.  
In the first step we construct the complex orthogonal matrix 
$\phi $ which diagonalizes $H^{PP}$. Here the columns of the 
matrix $\phi $ span the reference space $P$, and is a more 
convenient basis for perturbation expansions.  

Secondly, we perform a standard perturbation expansion in energy, and 
define $ M = \phi^T H^{PQ}$  which gives the orthogonal transformation of the
coupling block with respect to the reference states $\phi $. 
Using intermediate normalization, the energy corrections up to third order for 
a given state $\phi_i$ in the reference space,
may then be shown to be, 
\begin{eqnarray}
  E_i^0 &  = & \phi_i^T H^{PP} \phi_i, \:\: E_i^1 = 0, \\
  E_i^2 & = &  \sum_{j = N_P + 1}^N {M_{i,j}^2\over E_i^0 - H^0_{j,j} }, \\
  E_i^3 & = &  \sum_{j,k = N_P +1}^N  {M_{i,j} H^{QQ}_{j,k} M_{k,i}^T\over 
    ( E_i^0 - H^0_{j,j})( E_i^0 - H^0_{k,k}) }\: j \neq k, 
\end{eqnarray}
here $M_{i,j}$ is the dot product of the vector $\phi_i^T$ with column $j$ 
of the coupling block $H^{PQ}$.
Observe that there is no first order correction in the energy, meaning that
it has been accounted for by the reference states and energies. 

In the application of the multi-reference perturbation method to the calculation 
of multi-particle resonances
in Gamow shell-model calculations, we have to define a multi-particle model space which
describes most of the many-body correlations. 
Based on our knowledge from the two-particle system, 
we assume a  good choice for the   complement space
to consist of  configurations where all particles move in complex continuum states.
One would expect that the most important configurations of the multi-particle
resonance, are configurations which include single-particle resonances. 
In our calculation of $^7$He, the three-particle model space is chosen by studying the
squared amplitudes of the three-particle configurations given in Table~\ref{tab:tab11}.
There we see that the amplitudes of configurations where all particles move in continuum states
are small. In Refs.~\cite{witek1} and \cite{witek2}, where the Helium isotopes 
$^{6-9}$He were studied within the Gamow shell-model formulation, the authors reached similar 
conclusions.  Note also that we use the bare two-body interaction of Eq.~(\ref{eq:twobody}). 
This is to be contrasted to the method outlined in the next section where we use the Lee-Suzuki
transformation in order to define an effective interaction.

\begin{figure}[hbtp]
\begin{center}
\resizebox{8cm}{6cm}{\epsfig{file=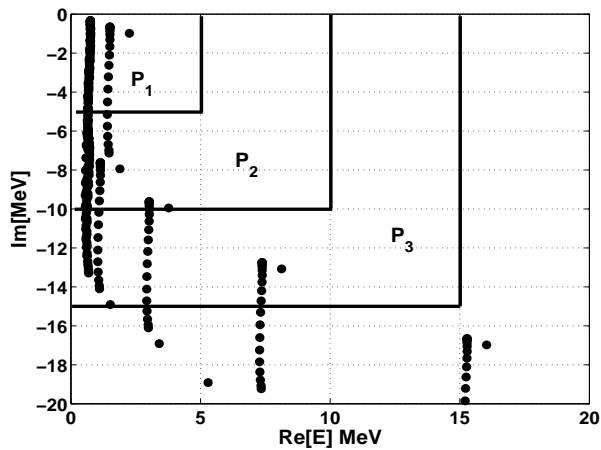}}
\end{center}
\caption{Three choices of the model space used in the multi-configuration 
perturbation method. The three-particle model space states are constructed
such that at most two particles move in the non-resonant continuum.} 
\label{fig:P_space_ch1}
\end{figure}

The three-particle model space, and corresponding complement space, 
used in our calculations is defined by
\begin{eqnarray}
  P \equiv 
  \left\{ \begin{array}{c} \vert RRR \rangle, \vert RRC\rangle, \vert RCC\rangle ,\\
    \mathrm{Re}(e_a+e_b+e_c) <  E_{\mathrm{cut}}, \\
    \mathrm{Im}(e_a+e_b+e_c) > -E_{\mathrm{cut}}
  \end{array} \right\}\:\:\: Q = 1-P,
  \label{eq:mspace1}
\end{eqnarray}
where at most two particles move in continuum states. We have also introduced a 
rectangular cutoff in the complex energy plane, 
since we assume that three-particle configurations high in 
the energy play a minor role on the formation of low-lying resonances.
Fig.~\ref{fig:P_space_ch1} gives a plot of the unperturbed three-particle 
spectrum where at most two particles move in complex continuum states, 
and three different cut-offs in energies and corresponding model spaces are shown.
 
\begin{figure}[hbtp]
\begin{center}
\resizebox{8cm}{6cm}{\epsfig{file=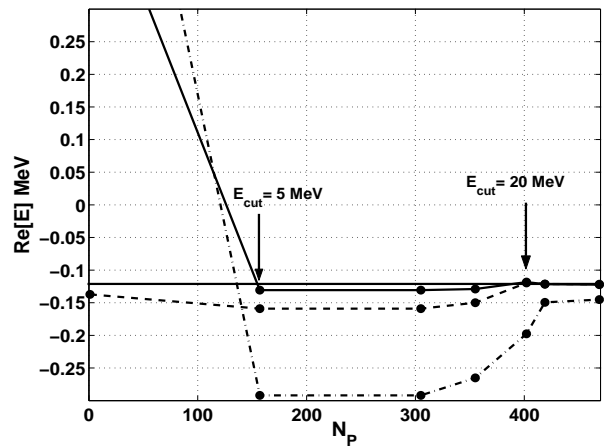}}
\end{center}
\caption{Convergence of the real part of the $J^{\pi}={3/2}^-$ resonance in  $^7$He, 
as the dimension of three-particle 
model space increases with increasing cutoff in energy. The cutoff in energy is
increased in steps of $5$ MeV, i.e. $E_{\mathrm{cut}} = 0,5,...,30$, and
given by the filled circles. The horizontal line is the
real part of the $J^{\pi}={3/2}^-$ resonance located at E $= -(0.120731 +0.122211i)$ MeV.
The dashed-dotted line is the zeroth order energy, the dashed line represents the second-order
energy and the solid line is the third-order energy.} 
\label{fig:multi_pert1}
\end{figure} 

\begin{figure}[hbtp]
\begin{center}
\resizebox{8cm}{6cm}{\epsfig{file=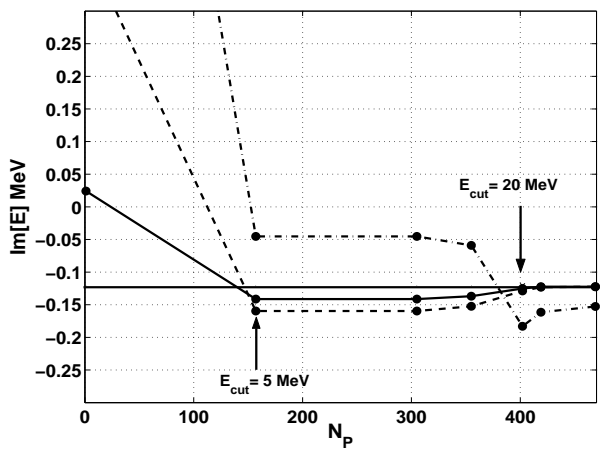}}
\end{center}
\caption{Convergence of the imaginary part of the $J^{\pi}={3/2}^-$ resonance in  $^7$He, 
as the number of three-particle 
model space increases with increasing cutoff in energy. 
The cutoff in energy is
increased in steps of $5$ MeV, i.e. $E_{\mathrm{cut}} = 0,5,...,30$, and
given by the filled circles.
The horizontal line is the
the imaginary part of the $J^{\pi}={3/2}^-$ resonance located at E $= -(0.120731 +0.122211i)$ MeV.
The dashed-dotted line is the zeroth order energy, the dashed line represents the second-order
energy while the solid line is the third-order energy. } 
\label{fig:multi_pert2}
\end{figure} 

Figs.~\ref{fig:multi_pert1} and ~\ref{fig:multi_pert2} show the convergence of the
real and imaginary part of the three-particle resonance energy in the 
multi-reference perturbation method up to third order. 
The model space used here is given in Eq.~\ref{eq:mspace1}, 
and the calculations were done for the increasing cutoffs in energy, $E_{\mathrm{cut}}=0,5,...,30$ MeV. 
Here the convergence is plotted with respect to the number of three-particle model-space
states $N_P$ for each energy cutoff. 
We see that a satisfactory convergence is obtained with $N_P \approx 400 $, 
corresponding to the energy cutoff $ E_{\mathrm{cut}} = 20$ MeV. 
As expected, we see that 
excitations of model space configurations located above $E_{\mathrm{cut}} \approx 5$ MeV yield
small contributions to the second- and third-order corrections to the resonance energy. 
Observe that the second- and third-order terms converge at the same number of model space states,
which indicates that second-order corrections in energy are seemingly 
sufficient for our applications. This is also an advantage from a numerical point of view. In second 
order one has to store only the diagonal part of the block $H^{QQ}$, while in third order
the complete block $H^{QQ} $ has be stored, which may be extremely large in many cases.
The zeroth order energy, which corresponds to diagonalization within $P$, does not 
saturate at the exact resonance energy with increasing $N_P$, which again shows that 
possible couplings with the $Q-$ space have to be accounted for, if one aims at accurate 
calculations.  

Summing up these results, we see that we obtain stable results with approximately
$N_P \approx 400 $ three-body configurations within the multi-reference perturbation method,
while the similarity transformation method of Sec.~\ref{sec:suzuki}
gives stable results for $N_P \approx 800 $ three-body configurations
for the same problem. The question now is whether we can marry these two approaches in our 
quest for smaller
Gamow shell-model spaces. This is the topic of the next section. 

\section{Effective Interaction Scheme for the Gamow Shell Model}
\label{sec:scheme}

In the previous sections it was shown that the Lee-Suzuki similarity transformation 
and the multi-reference perturbation method may be used in the Gamow shell model 
in order to  account for the most important correlations of for example a  multi-particle resonance. 
Although the dimensionality of the problem derived either 
from the similarity transformation method or the multi-reference perturbation method 
was small compared to the full problem, the dimensionality may still be a severe
problem when dealing with more than three particles in a big valence space. 

The drawback of the multi-reference perturbation method is that one has to store 
extremely large matrices $H^{QQ}$ 
if one wishes to go beyond second order in perturbation theory. In the similarity 
transformation method one does not have to deal with $H^{QQ}$, as couplings with
the $Q$-space states have been dealt with, in practical calculations at least at the two-body level. 
Going to systems with larger degrees of freedom, the $P$-space may nevertheless,
at the converged level, be too large for our brute force diagonalization approach.

The aim of this section is to propose an effective interaction and perturbation theory scheme 
for the Gamow shell model. This approach 
combines the similarity transformation method and the multi-reference perturbation method,
so that hopefully multi-particle resonances where several particles move 
in large valence spaces, may be calculated without a diagonalization in the full space.
Our algorithm is as follows
\begin{enumerate}
\item Choose an optimal set of $n_{sp}$ single-particle orbits, which in turn define 
two-body  $P_{2p}$ and  many-body spaces. In our case  these single-particle orbits are defined by 
selected states in $^5$He.
\item Construct a two-particle effective interaction by the Lee-Suzuki similarity transformation method
  within the two-particle model space $P_{2p}$. Such diagonalizations can be done for very large spaces,
see for example Refs.~\cite{bruce1,bruce2,bruce3,bruce4}.
\item The next step is to divide the multi-particle model space $P$ in two smaller spaces $P'$ and $Q'$, 
  where $P=P'+Q'$ and $ N_P = N_{P'} +N_{Q'}  $. 
  The choice of $P'$ should be dictated by our knowledge of the physical system. As an example, 
one may consider those single-particle configurations within the $P-$space that play 
the dominant role in the formation of 
  the multi-particle resonance.  The number $N_P = N_{P'} + N_{Q'}$ represents the total number of 
  many-body configurations within the $P-$space.
\item Now that we have divided the $P$-space in two sub-spaces $P'$ and $Q'$, we use for example 
  the multi-reference perturbation method  to account for 
  excitations from the $P'-$ space to the $Q'-$ space to obtain energy corrections to a specific order.
  Increase the size of the $P'-$space until convergence is obtained. In the case $N_{P'} = N_P $ and 
  $N_{Q'} = N_P-N_{P'} = 0$ the multi-reference perturbation expansion terminates at zeroth order,
  and corresponds to a full diagonalization within the $P-$space. 
  Another option is to use for example the coupled cluster method as exposed in Refs.~\cite{cc1,cc2}. 
\item Start from top again with a larger set of single-particle orbits, 
  and continue until a  convergence criterion is reached.
\end{enumerate}
We illustrate these various choices of model spaces in the following two figures.
Fig.~\ref{fig:paulioperator} defines our model space for the Lee-Suzuki similarity transformation
at the two-body level. This corresponds to steps one and two in the above algorithm.
The set of single-particle orbits defines the last single-particle orbit in the model space $n_{sp}$. 
Note that we could have chosen a model space defined by a cut in  energy, as done by the No-Core
collaboration, see for example Refs.~\cite{bruce1,bruce2,bruce3,bruce4}. These examples serve just to 
illustrate the algorithm.
\begin{figure}[htb]
\begin{minipage}[t]{80mm}
\setlength{\unitlength}{0.6cm}
\begin{picture}(9,10)
\thicklines
   \put(1,0.5){\makebox(0,0)[bl]{
              \put(0,1){\vector(1,0){8}}
              \put(0,1){\vector(0,1){8}}
              \put(-0.6,6){\makebox(0,0){$n_{sp}$}}
              \put(5,0.5){\makebox(0,0){$n_{sp}$}}
              \put(2,3){\makebox(0,0){$\hspace{1cm}P_{2p}$}}
              \put(2,7){\makebox(0,0){$\hspace{1cm}Q_{2p}=1-P_{2p}$}}
              \put(0,6){\line(1,0){5}}
              \put(5,1){\line(0,1){5}}
         }}
\end{picture}
\caption{Possible definition of the two-body 
exclusion operator $Q_{2p}=1-P_{2p}$ used to compute the 
Lee-Suzuki similarity transformation and its effective interaction at the two-body level. 
The border of the model space is defined by the last single-particle orbit $n_{sp}$. 
\label{fig:paulioperator}}
\end{minipage}
\hspace{\fill}
\begin{minipage}[t]{75mm}
\setlength{\unitlength}{0.6cm}
\begin{picture}(9,10)
\thicklines
   \put(1,0.5){\makebox(0,0)[bl]{
              \put(0,1){\vector(1,0){8}}
              \put(0,1){\vector(0,1){8}}
              \put(-0.6,6){\makebox(0,0){$N_P$}}
              \put(5,0.5){\makebox(0,0){$N_P$}}
              \put(-0.6,3){\makebox(0,0){$N_{P'}$}}
              \put(2,0.5){\makebox(0,0){$N_{P'}$}}
              \put(0,6){\line(1,0){5}}
              \put(5,1){\line(0,1){5}}
              \put(0,3){\line(1,0){2}}
              \put(2,1){\line(0,1){2}}
         }}
\end{picture}
\caption{Possible definition of many-body space $N_P$ and reduced space $N_P'$. \label{fig:finalp}}
\end{minipage}
\end{figure}
Fig.~\ref{fig:finalp} demonstrates again a possible division of the space into the full model space
$N_P$ and a smaller space $N_{P'}$. Again, this 
figure serves only the purpose of illustrating the method.
In our actual calculations we define the smaller space $N_{P'}$ via an energy cut in the real and 
imaginary eigenvalues and selected many-body configurations.
 
In summary, defining a set of single-particle orbits in order to construct the two-body and many-body 
model spaces, we obtain first an effective two-body interaction in the  space $P_{2p}$
by performing the 
Lee-Suzuki \cite{suzuki1,suzuki2,suzuki3,suzuki4} transformation. This interaction and the 
pertinent single-particle orbits are then used to define a large many-body space. 
It is therefore of interest to see if we can reduce this dimensionality through the definition of smaller
spaces and perturbative corrections.

We present here as a test case, the calculation of the $J^{\pi}=3/2^-$ three-particle resonance
within the perturbation scheme outlined above. 
$24$ single-particle orbits for the $lj$ configuration $p_{3/2}$ are included, giving a total 
dimension of $d = 9224$ for the $J={3/2}$ three-particle basis.

We define five different model spaces $P$, 
given by the total number of three-body configurations
$N_P$. The number of single-particle orbits and three-body states are listed 
in Table.~\ref{tab:tab12}. The single-particle model space, defining 
$P$, is constructed according to the prescription outlined in section~\ref{sec:suzuki}. 
The reference space $P'$ which defines a $proper subset$ of each model space $P$, is again 
defined by 
\begin{eqnarray}
  {P'}_{i} \equiv 
  \left\{ \begin{array}{c} \vert RRR \rangle, \vert RRC\rangle, \vert RCC\rangle \\
    \mathrm{Re}(e_a+e_b+e_c) <  E_{\mathrm{cut}} \\
    \mathrm{Im}(e_a+e_b+e_c) > -E_{\mathrm{cut}}   
  \end{array} \right\} \:\: \subset P_i.
  \label{eq:mspace3}
\end{eqnarray}
\begin{table}[htbp]
\caption{Five different $P-$spaces defined for 
increasing number of single-particle model space orbits $n_{sp}$
consisting of the $lj$ configuration $p_{3/2}$.
The number $N_P$ gives the dimension of the three-particle model space $P$
for $J^{\pi}=3/2^-$ with a full dimensionality with  $n_{sp}=24$ of 
$N_P=9224$.}
\begin{tabular}{cccccc}\hline
  \hline
  $P$ & $ P_1$ & $P_2$ & $P_3 $ & $P_4$ & $P_5$  \\
  \hline
  $n_{sp}$ & 8 & 10 & 12 & 14 & 16 \\
  \hline
  $N_P $ & 344 & 670 & 1156 & 1834 & 2736 \\
  \hline 
\end{tabular}
\label{tab:tab12}
\end{table}
The reader should note that for each space $P_1$, $P_2$ and so forth 
listed in Table \ref{tab:tab12}, we can compare the results from this perturbative analysis
with those from the exact diagonalization done in these spaces. 
This is shown in Tables~\ref{tab:tab13},\ref{tab:tab14}, \ref{tab:tab15}, \ref{tab:tab16} and \ref{tab:tab17}.

\begin{table}[htbp]
  \caption{ Resonance energy to second (E$^2$) and third order (E$^3$) in the multi-reference 
    perturbation expansion, for the model space $P_1$ given in Table~ \ref{tab:tab12}. 
    The subspaces ${P'}_1$ are defined for 
    different energy cutoffs, increased in steps of $10$ MeV. 
    In the last line we give the exact energy within $P_1$.
    Energies are given in units of MeV.}
  \begin{tabular}{ccccccc}
    \hline
    \multicolumn{1}{c}{} & 
    \multicolumn{3}{c}{$N_{P_1}= 344 $} & 
    \multicolumn{3}{c}{$N_{{P'_1}_{\mathrm{max}}} = 113 $}\\
    \hline
    \multicolumn{1}{c}{$N_{{P'}_1}$} & \multicolumn{1}{c}{$N_{Q'_1}$} & 
    \multicolumn{1}{c}{$E_{\mathrm{cut}}$} & 
    \multicolumn{1}{c}{Re[E$^2$] }&\multicolumn{1}{c}{Im[E$^2$]} &
    \multicolumn{1}{c}{Re[E$^3$] }&\multicolumn{1}{c}{Im[E$^3$]} \\
    \hline
    1   & 343 & 0  & 0.066 &  0.322 &  0.606 &  0.088 \\
    113 & 231 & 10 & 0.041 & -0.075 & 0.041 & -0.076 \\
    \hline
    \multicolumn{3}{c}{Exact within $P_1$:} & 0.042 & -0.076 \\
    \hline
  \end{tabular}
  \label{tab:tab13}
\end{table}

\begin{table}[htbp]
  \caption{ Resonance energy to second (E$^2$) and third order (E$^3$) in the multi-reference 
    perturbation expansion, for the model space $P_2$ given in Table~ \ref{tab:tab12}. 
    The subspaces ${P'}_2$ are defined for 
    different energy cutoffs, increased in steps of $10$ MeV. 
    In the last line we give the exact energy within $P_2$.
    Energies are given in units of MeV.}
  \begin{tabular}{ccccccc}
    \hline
    \multicolumn{1}{c}{} & 
    \multicolumn{3}{c}{$N_{P_2}= 670 $} & 
    \multicolumn{3}{c}{$N_{{P'_2}_{\mathrm{max}}} = 181 $}\\
    \hline
    \multicolumn{1}{c}{$N_{P'_2}$} & \multicolumn{1}{c}{$N_{Q'_2}$} & 
    \multicolumn{1}{c}{$E_{\mathrm{cut}}$} & 
    \multicolumn{1}{c}{Re[E$^2$] }&\multicolumn{1}{c}{Im[E$^2$]} &
    \multicolumn{1}{c}{Re[E$^3$] }&\multicolumn{1}{c}{Im[E$^3$]} \\
    \hline
    1   & 669 & 0  & -0.053 &  0.357 & 0.562 &  0.059 \\
    157 & 513 & 10 & -0.078 & -0.110 & -0.079 &  -0.110 \\
    181 & 489 & 20 & -0.082 & -0.110 & -0.083 &  -0.110 \\
    \hline
    \multicolumn{3}{c}{Exact within $P_2$:} & -0.081 &  -0.110 \\
    \hline
  \end{tabular}
  \label{tab:tab14}
\end{table}

\begin{table}[htbp]
  \caption{ Resonance energy to second (E$^2$) and third order (E$^3$) in the multi-reference 
    perturbation expansion, for the model space $P_3$ given in Table~ \ref{tab:tab12}. 
    The subspaces ${P'}_3$ are defined for 
    different energy cutoffs, increased in steps of $10$ MeV. 
    In the last line we give the exact energy within $P_3$.
    Energies are given in units of MeV.}
  \begin{tabular}{ccccccc}
    \hline
    \multicolumn{1}{c}{} & 
    \multicolumn{3}{c}{$N_{P_3}= 1156 $} & 
    \multicolumn{3}{c}{$N_{{P'_3}_{\mathrm{max}}} = 265 $}\\
    \hline
    \multicolumn{1}{c}{$N_{P'_3}$} & \multicolumn{1}{c}{$N_{Q'_3}$} & 
    \multicolumn{1}{c}{$E_{\mathrm{cut}}$} & 
    \multicolumn{1}{c}{Re[E$^2$] }&\multicolumn{1}{c}{Im[E$^2$]}&
    \multicolumn{1}{c}{Re[E$^3$] }&\multicolumn{1}{c}{Im[E$^3$]} \\
    \hline
    1   & 1155 & 0 & -0.099 &  0.378 & 0.561 &  0.050 \\ 
    205 & 951 & 10 & -0.114 & -0.134 & -0.114 & -0.133 \\
    265 & 891 & 20 & -0.117 & -0.130 & -0.118 & -0.130 \\
    \hline
    \multicolumn{3}{c}{Exact within $P_3$:} &  -0.116 & -0.130 \\
    \hline
  \end{tabular}
  \label{tab:tab15}
\end{table}

\begin{table}[htbp]
  \caption{ Resonance energy to second (E$^2$) and third order (E$^3$) in the multi-reference 
    perturbation expansion, for the model space $P_4$ given in Table~ \ref{tab:tab12}. 
    The subspaces ${P'}_4$ are defined for 
    different energy cutoffs, increased in steps of $10$ MeV. 
    In the last line we give the exact energy within $P_4$.
    Energies are given in units of MeV.}
  \begin{tabular}{ccccccc}
    \hline
    \multicolumn{1}{c}{} & 
    \multicolumn{3}{c}{$N_{P_4}= 1834 $} & 
    \multicolumn{3}{c}{$N_{{P'_4}_{\mathrm{max}}} = 365 $}\\
    \hline
    \multicolumn{1}{c}{$N_{P'_4}$} & \multicolumn{1}{c}{$N_{Q'_4}$} & 
    \multicolumn{1}{c}{$E_{\mathrm{cut}}$} &
    \multicolumn{1}{c}{Re[E$^2$] }&\multicolumn{1}{c}{Im[E$^2$]}&
    \multicolumn{1}{c}{Re[E$^3$] }&\multicolumn{1}{c}{Im[E$^3$]} \\
    \hline
    1   & 1833 & 0  & -0.134 &  0.397  & 0.532 &  0.026 \\
    253 & 1581 & 10 & -0.155 & -0.160  & -0.130 & -0.141 \\
    347 & 1487 & 20 & -0.119 & -0.127  & -0.120 & -0.126 \\
    365 & 1469 & 30 & -0.122 & -0.123  & -0.123 & -0.124 \\
    \hline
    \multicolumn{3}{c}{Exact within $P_4$:} & -0.121 & -0.124 \\
    \hline
  \end{tabular}
  \label{tab:tab16}
\end{table}

\begin{table}[htbp]
  \caption{ Resonance energy to second (E$^2$) and third order (E$^3$) in the multi-reference 
    perturbation expansion, for the model space $P_5$ given in Table~ \ref{tab:tab12}. 
    The subspaces ${P'}_5$ are defined for 
    different energy cutoffs, increased in steps of $10$ MeV. 
    In the last line we give the exact energy within $P_5$.
    Energies are given in units of MeV.}
  \begin{tabular}{ccccccc}
    \hline
    \multicolumn{1}{c}{} & 
    \multicolumn{3}{c}{$N_{P_5}= 2736 $} & 
    \multicolumn{3}{c}{$N_{{P'_5}_{\mathrm{max}}} = 419 $}\\
    \hline
    \multicolumn{1}{c}{$N_{P'_5}$} & \multicolumn{1}{c}{$N_{Q'_5}$} & 
    \multicolumn{1}{c}{$E_{\mathrm{cut}}$} & 
    \multicolumn{1}{c}{Re[E$^2$] }&\multicolumn{1}{c}{Im[E$^2$]}&
    \multicolumn{1}{c}{Re[E$^3$] }&\multicolumn{1}{c}{Im[E$^3$]} \\
    \hline
    1   & 2735 & 0  & -0.137 &  0.399 & 0.530 &  0.024 \\
    253 & 2483 & 10 & -0.159 & -0.160 & -0.131 & -0.141 \\
    347 & 2389 & 20 & -0.120 & -0.129 & -0.118 & -0.125 \\
    409 & 2327 & 30 & -0.122 & -0.122 & -0.122 & -0.122 \\
    419 & 2317 & 40 & -0.122 & -0.122 & -0.123 & -0.122 \\
    \hline
    \multicolumn{3}{c}{Exact within $P_5$:} & -0.121 & -0.122 \\
    \hline
  \end{tabular}
  \label{tab:tab17}
\end{table}

As the number of reference states $N_{P'}$ increases with increasing cutoff in energy $E_{\mathrm{cut}}$, 
one reaches a maximum of reference states $N_{ {P'}_{\mathrm{max}}}$ within each $P-$space. 
From the definition of the reference space $P'$ in Eq.~(\ref{eq:mspace3}), 
it will never coincide with the $P-$space as 
one exhausts the number of configurations $ \vert RRR \rangle, \vert RRC\rangle, \vert RCC\rangle $ 
within $P$, since one by definition never includes the configurations $ \vert CCC \rangle$, 
i.e. $\{ \vert P' \rangle \} \subset \{ \vert P \rangle \}$.
The perturbation scheme for a reference space $P'$ given by  Eq.~(\ref{eq:mspace3}), 
will therefore only yield convergent results as long as our assumption 
that the configurations $\vert CCC\rangle $ play a minor role compared to
the reference states. Although the configurations $\vert CCC \rangle $ 
turn out to play a minor role for the states we have considered in this work, 
there is no a priori  reason for this to be the case when considering other multi-particle resonances.
If no convergence is observed, one should simply choose another reference space $P'$, 
based for example on the single-particle model space, see Figs.~\ref{fig:paulioperator} and ~\ref{fig:finalp}. 

\begin{figure}[hbtp]
\begin{center}
\resizebox{8cm}{6cm}{\epsfig{file=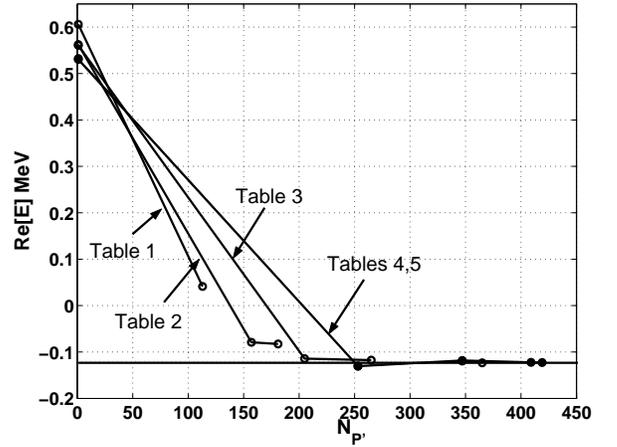}}
\end{center}
\caption{Convergence of the real part of the $J^{\pi}=3/2^-$ resonance energy 
in $^7$He within the perturbative scheme outlined in the text, for the different 
model spaces $P$ given in Table~\ref{tab:tab12}. The open circles along the different
solid lines gives the calculations within each $P_i$.  $N_{P'}$ gives the number
of reference states in $P'$, which is a subspace of $P$.
The horizontal line is the
the real part of the $J^{\pi}={3/2}^-$ resonance located at E $= -(0.120731 +0.122211i)$ MeV.} 
\label{fig:he7_pert3}
\end{figure}

\begin{figure}[hbtp]
\begin{center}
\resizebox{8cm}{6cm}{\epsfig{file=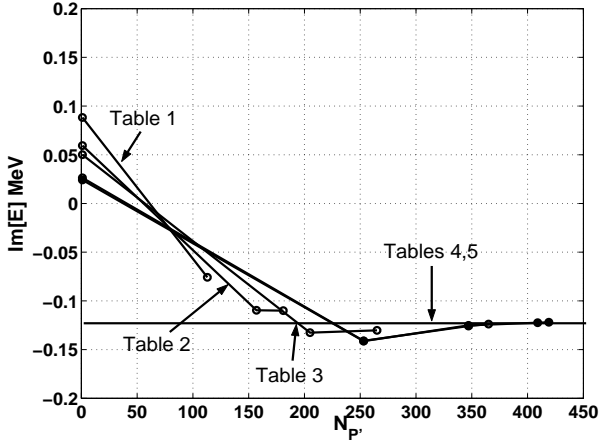}}
\end{center}
\caption{Convergence of the imaginary part of the $J^{\pi}=3/2^-$ resonance energy 
in $^7$He within the perturbative scheme outlined in the text, for the different 
model spaces $P$ given in Table~\ref{tab:tab12}. The open circles along the different
solid lines gives the calculations within each $P_i$.  $N_{P'}$ gives the number
of reference states in $P'$, which is a subspace of $P$.
The horizontal line is the
the imaginary part of the $J^{\pi}={3/2}^-$ resonance located at E $= -(0.120731 +0.122211i)$ MeV.} 
\label{fig:he7_pert4}
\end{figure}

Figs.~\ref{fig:he7_pert3} and \ref{fig:he7_pert4} gives plots of the real and imaginary 
part of the resonance energy to third order in the multi-reference perturbation expansion for 
the different model spaces considered above. From the plot one concludes that convergence 
is obtained for a small number of reference states $N_{P'}\sim 350-400$.  

In the approach considered above, the dimension of the $Q'-$space is considerably smaller 
than the dimension of the complement space $Q = 1-P$, which makes it much less time and memory 
consuming to compute the matrix elements of $H^{Q'Q'}$. 
We have seen from the above calculations that a termination of the perturbation expansion
at second order compares well with the rate of convergence for the third-order expansion.
This makes it numerically feasible to treat systems where several particles 
move in a large valence space, within perturbative scheme outlined above.

We conclude this work by applying 
our scheme to the calculation 
of the three-particle resonances in $^7$He, where $24$ single particle states
for each of the $lj$ single-particle states $p_{1/2}$ and $p_{3/2}$ are included. 
The Hamiltonian for the $J^{\pi}=1/2^-,3/2^-$ and $ J^{\pi}=5/2^-$ 
states for $^7$He has dimensions $N_P = 29648 , 38896$ and 
$N_P = 27072 $, respectively. 
The main component of the resonant wave functions 
turns out to be the $\vert RRR \rangle $ configuration, which gives the allowed couplings  
and corresponding unperturbed energies and wave functions
$\vert \left( p_{3/2} \right)^3; J^\pi = {3/2}^-_1 \rangle$ with energy
$E_0 =  \left(2.2560 - 0.9840i\right)$ MeV, 
$\vert  p_{1/2}\:\left( p_{3/2}\right)^2_0 ;\: J^\pi = {1/2}^-_1 \:\rangle$ with energy 
$E_0 = \left(3.6581 - 3.5681i\right)$ MeV,
$\vert  p_{1/2}\:\left( p_{3/2}\right)^2_2 ;\: J^\pi = {3/2}^-_2, \:{5/2}^-_1 \:\rangle$ with energy
$ E_0 = \left(3.6581 - 3.5681i\right)$ MeV and
$\vert \left( p_{1/2} \right) ^2_0 \: p_{3/2} ; J^\pi = {3/2}^-_3 \rangle$ and energy
$E_0 =  \left( 5.0602 - 6.1522i\right)$.
We report here only the converged results for the lowest-lying $^7$He resonances. They are
$E({3/2}_1^-) = (0.02-0.08i )$ MeV, 
$E({1/2}_1^-) = (0.39 -3.98i )$ MeV,
$E({3/2}_2^-) = (2.43 -1.95i)$ MeV, 
$E({5/2}_1^-) = (2.75-0.89i)$MeV, and 
$E({3/2}_3^-) = (3.85 -3.06i)$ MeV. Using our combination of the Lee-Suzuki  similarity transformation and the 
multi-reference perturbation method, results close to the exact ones where obtained with approximately
$N_{P'} \sim 1400-2000$ three-particle configurations. This is a considerable reduction compared
with the full dimensionalities listed above.

\begin{figure}[hbtp]
\begin{center}
\resizebox{8cm}{6cm}{\epsfig{file=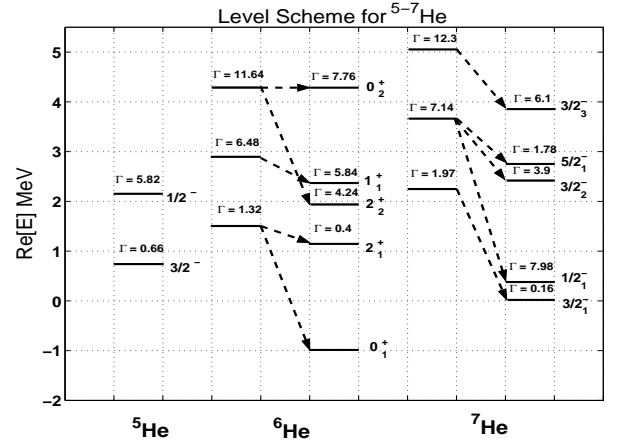}}
\caption{Energy levels 
for the nuclei $^{5,6,7}$He for model space consisting of the 
$lj$ single-particle states $p_{1/2}$ and $p_{3/2}$. The unperturbed energy levels 
for $^6$He and $^7$He are shown to the left in each case.}
\label{fig:level_scheme} 
\end{center}
\end{figure}

Fig.~\ref{fig:level_scheme} displays the calculated energy levels 
for the nuclei $^{5-7}$He within our model. The unperturbed energy levels 
for $^6$He and $^7$He are also shown, and serve to illustrate how the 
two- and three-particle resonances develop when the nucleon-nucleon interaction
from Eq.~(\ref{eq:twobody})  is included. 
There are several interesting features which can be seen from Fig.~\ref{fig:level_scheme}. 
The $0^+$- and $2^+$-states in $^6$He are formed within our model
due to a strong pairing effect between the two neutrons moving in equivalent orbits.
When the nucleon-nucleon interaction is included, we observe that the $0^+$ and $2^+$ unperturbed 
energies gain an additional attraction.
On the other hand, the $1^+$ state in $^6$He exhibits weak pairing effects, 
the main component of the $1^+$ wave function is still the configuration $\vert RR\rangle $,
but here the dominant contribution comes from a coupling between the 
$p_{1/2}$ and $p_{3/2}$ single-particle resonances in $^5$He. 
Adding the nucleon-nucleon interaction,
does not change the unperturbed energy level significantly, see Fig.~\ref{fig:level_scheme}.
The calculations done within our crude model of $^7$He, may suggest that this unbound 
nucleus has an even richer continuum structure than proposed in the recent review of Jonson, see
Ref~\cite{jonson}.  
In Ref.~\cite{jonson} two excited states with tentative spins $J^\pi = {1/2}^- $ and $J^\pi = {5/2}^- $
where reported to exist above the ground state in $^7$He. The level spacings 
relative to the $^7$He ground state were reported to be $0.57$ MeV and $2.87$ MeV  for the 
$J^\pi = {1/2}^- $ and $J^\pi = {5/2}^-$ states, respectively. The main decay channel of the 
$J^\pi = {5/2}^-$ resonance at $2.87$ Mev
is $\alpha + 3n$. From this decay channel,  Jonson Ref.~\cite{jonson} concluded
that the configuration $\vert  p_{1/2}\:\left( p_{3/2}\right)^2_2 ; J^\pi = {5/2}^- \rangle $ is
the most probable one.
In our calculations we find a rich continuum structure in the energy region $\mathrm{Re}[E] \approx 3$ MeV
above threshold. 
From Fig.~\ref{fig:level_scheme} it is seen that the resonant three-particle states 
$J^{\pi} = {3/2}^+_2, {5/2}^+_2 $ and $J^{\pi} = {3/2}^+_3$ are located rather close in real energy.
Although the widths vary from $2-6$ MeV, this observation 
raises the question of whether these structures may be observed, and which
spin assignements and the nature of the experimentally observed 
structures around $2.87$ MeV in $^7$He are.
Further it is seen that the real part of the $J^\pi = {1/2}^-$ resonance changes strongly, 
and moves towards the threshold when the nucleon-nucleon interaction is included.
However, these results must be gauged with the fact that we are using a purely phenomenological
nucleon-nucleon interaction model. The inclusion of a realistic interaction is the topic for a future work.
The main issue here was to 
demonstrate how to derive effective interactions for the Gamow shell model, with a
considerable reduction in dimensionality.

\section{Conclusion and future perspectives.}
\label{sec:conclusion}

In this work we have applied the contour deformation method in momentum space, with  
a single-particle basis in momentum space serving as a starting point for 
Gamow shell-model calculations. 
The resonant spectra of the drip-line nuclei $^{5-7}$He have been studied
and described using phenomenologically derived nucleon-nucleon interactions. 
It was illustrated that the choice of contour gives a good convergence for various resonant states, 
and in addition allows for a clear distinction between all physical states and the
remaining complex continuum states. The main purpose of this work was to propose an 
effective interaction scheme for Gamow shell-model calculations. One of the most
severe difficulties regarding Gamow shell-model 
calculations is the dramatic growth of the shell-model 
dimension when dealing with several valence particles moving in a large shell-model
space. This dimensionality problem is even more severe 
than the harmonic oscillator 
representation used in traditional shell-model equation studies,
In the Berggren representation a large number of complex-continuum 
states has to be included as well. The clear distinction of the unperturbed 
resonances from the dense distribution of complex continuum states, allows for 
a perturbation treatment, when configuration mixing is taken into account. 
For perturbation
expansions to converge, the unperturbed states have to be well 
separated from the $Q$-space
states, or else the propagators will contain poles which make a  perturbative treatment 
difficult. It has been shown that the Lee-Suzuki similarity 
transformation combined with the
multi-reference perturbation method, reduces the full problem to about $3-4\%$. 

Treating the many-particle problem in some perturbation scheme, we need to define
a reference (model) space which describes most of the many-body correlations. The
method and scheme outlined here, allows for a perturbative treatment of many-body 
states in which anti-bound states may play an important role, such as in the drip-line
nuclei $^{11}$Li. 

As has been pointed out, the location of multi-particle resonances depends on the
effective interaction used between valence nucleons. The next step 
is to derive a realistic effective interaction for Gamow shell-model 
calculations, and
self-consistent Hartree-Fock single-particle energies for loosely bound 
nuclei, starting from a realistic nucleon-nucleon force. Using the Berggren 
representation may give an underlying understanding of many-body resonances 
from a microscopic point of view. 
Moreover, in our algorithm of Sec.~\ref{sec:scheme} we employed the multi-reference
perturbation method. Our future plans involve replacing this method by the Coupled
Cluster approaches, as discussed in Refs.~\cite{cc1,cc2}.

\section*{Acknowledgments}
Support by the Research Council of Norway is greatly acknowledged.

\end{document}